\documentclass[amsmath,amssymb,aps,pre,reprint,superscriptaddress]{revtex4-1}
\usepackage{graphicx}
\usepackage{bm}

\begin{document}

\title{Plastic systemic inhibition controls amplitude while allowing phase pattern in a stochastic neural field model}


\author{Conor L. Morrison}
\affiliation{Department of Statistics, University of British Columbia, Vancouver, BC Canada V6T 1Z4}
\author{Priscilla E. Greenwood}
\affiliation{Department of Mathematics, University of British Columbia, Vancouver, BC Canada V6T 1Z2}
\author{Lawrence M. Ward}
\affiliation{Department of Psychology and Djavad Mowafaghian Centre for Brain Health, 2136 West Mall, University of British Columbia, Vancouver, BC Canada V6T 1Z4}

\date{\today}

\begin{abstract}
Oscillatory phase pattern formation and amplitude control for a linearized stochastic neuron field model was investigated by simulating coupled stochastic processes defined by stochastic differential equations. It was found, for several choices of parameters, that pattern formation in the phases of these processes occurred if and only if the amplitudes were allowed to grow large. Stimulated by recent work on homeostatic inhibitory plasticity, we introduced static and plastic (adaptive) systemic inhibitory mechanisms to keep the amplitudes stochastically bounded in subsequent simulations. The systems with static systemic inhibition exhibited bounded amplitudes but no sustained phase patterns, whereas the systems with plastic systemic inhibition exhibited both bounded amplitudes and sustained phase patterns. These results demonstrate that plastic inhibitory mechanisms in neural field models can stochastically control amplitudes while allowing patterns of phase synchronization to develop. Similar mechanisms of plastic systemic inhibition could play a role in regulating oscillatory functioning in the brain.\end{abstract}


\maketitle

\section{Introduction}
\label{intro}
Mathematical models that exhibit oscillations have played a key role in modeling neural phenomena. Such models include models of individual neurons --- e.g., leaky integrate-and-fire neurons \cite{S2013}, the Izhikevich neuron \cite{I2008}, or the Hodgkin-Huxley neuron \cite{HH1952} --- and models of neuron populations, such as the Wilson-Cowan (W-C) model \cite{WC1972} or neural field models such as those studied by Faugeras and Inglis \cite{FI2015}. Perhaps the most influential model of neuron populations has been that of Wilson and Cowan \cite{WC1972}. They proposed a nonlinear rate model for two interacting populations of neurons:
\begin{equation}\label{WCE1}
\begin{split}
& dV_E(t)=\bigg[-V_E(t)+g\Big[a_E\Big(S_{EE}V_E(t)\\
&-S_{EI}V_I(t)-\theta_E+P_E(t)\Big)\Big]\bigg]dt,
\end{split}
\end{equation}
\begin{equation}\label{WCI1}
\begin{split}
& dV_I(t)=\bigg[-V_I(t)+g\Big[a_I\Big(-S_{II}V_I(t)\\
&+S_{IE}V_E(t)-\theta_I\Big)\Big]\bigg]dt,
\end{split}
\end{equation}
where $V_E(t), V_I(t)$ are voltages of excitatory and inhibitory neuron populations, respectively, $S_{EE}, S_{II}, S_{EI}, S_{IE}$ are synaptic efficacies that control their interactions, $g$ is a sigmoid threshold function, $a_E, a_I, \theta_E, \theta_I$ are constants, and $P_E(t)$ is input current. In the W-C model \eqref{WCE1}, \eqref{WCI1} and in most neural field models a nonlinear transform $g$ is introduced in order to bound the amplitudes, where $g$ is typically the logistic function or some other sigmoid function. This allows the formation of limit cycles that resemble oscillations recorded from neural systems. 

Nonlinear models of neuron populations like the W-C model can be linearized and centered at a relevant fixed point, resulting in systems like that studied by Kang et al. \cite{K2010} and by Greenwood et al. \cite{GMW2015}. The latter system, which we will employ in what follows, is governed by stochastic differential equations for mean voltages of excitatory and inhibitory neuron populations --- $E$ and $I$, respectively --- given by,
\begin{equation}\label{WCE}
\begin{split}
& \tau_E dE(t)=(-E(t)+S_{EE}E(t)\\
&-S_{EI}I(t))dt+\sigma_E  dW_E(t),
\end{split}
\end{equation}
and,
\begin{equation}\label{WCI}
\begin{split}
& \tau_I dI(t)=(-I(t)-S_{II}E(t)\\
&+S_{IE}E(t)) dt+ \sigma_I dW_I(t),
\end{split}
\end{equation}
or equivalently by the matrix equation,
\begin{equation}\begin{split} \label{eimatrix1}
d\mathbb{V}=[E(t),I(t)]^\top=-\mathbb{A} \mathbb{V}(t) dt+\mathbb{N} d\mathbb{W},
\end{split}\end{equation}
where,
\begin{equation}\begin{split}\label{matrixa}
-\mathbb{A}=
\begin{pmatrix}
(-1+S_{EE})\tau_E^{-1} & -S_{EI}\tau_E^{-1} \\
S_{IE}\tau_I^{-1} & (-1-S_{II})\tau_I^{-1}
\end{pmatrix},
\end{split}\end{equation}
where $d\mathbb{W}=(dW^E,dW^I)^\top$ are independent standard Brownian motions, and where,
\begin{equation}\begin{split}\label{nmat}
\mathbb{N}=
\begin{pmatrix}
\sigma_E\tau_E^{-1} & 0 \\
0 & \sigma_I\tau_I^{-1}
\end{pmatrix}.
\end{split}\end{equation} 
Here $\tau_E$ and $\tau_I$ represent time constants, $S_{II}$, $S_{IE}$, $S_{EI}$, and $S_{EE}$ represent mean synaptic connection efficacies, and $\sigma_I$ and $\sigma_E$ are amplitudes. We will refer to such a system as an EI-pair system.

The model \eqref{eimatrix1} is a form of W-C model but with the addition of time constants and stochastic terms and the omission of several other constants, the input current, and, most importantly for our purposes, the amplitude-bounding sigmoid function $g$ (as in \eqref{WCE1}, \eqref{WCI1}). Without the amplitude-bounding function, the model \eqref{eimatrix1} does not exhibit a noisy limit cycle, as a stochastic form of the W-C model does in \cite {Wallace2011}. In the absence of a limit cycle, however, oscillations can be sustained owing to the stochasticity in the system. These oscillations are referred to as `quasi-cycles' \cite{B2010} and occur when the matrix $-\mathbb{A}$, as in \eqref{eimatrix1}, has complex eigenvalues $-\lambda \pm \omega i$, with $0<\lambda$, and the matrix $\mathbb{N}\ne 0$. With $\mathbb{N}=0$ the system damps to the fixed point because the real part of the eigenvalues, $-\lambda$, is negative. Adding moderate noise sustains noisy oscillations at the `natural' frequency $\omega$. Quasi-cycles in Wilson-Cowan-type population models were first described by Bressloff \cite{B2010}, and those in systems such as \eqref{eimatrix1} were investigated mathematically by Baxendale and Greenwood \cite{BG2010}. 

Recent work has investigated \emph{coupled} systems of quasi-cycle oscillatory EI-pairs, a single pair of which whose stochasticity has been shown to reproduce qualitative behaviour observed in neural systems, such as gamma bursts \cite{GMW2015}. In \cite{GMW2016}, a Kuramoto-type coupling scheme was applied to the phases ($\theta(t)= \arctan(\frac{I(t)}{E(t)})$) and amplitudes ($Z(t)=||\mathbb{V}(t)||$) of EI-pairs under an approximation given by Baxendale and Greenwood in \cite{BG2010}. A linear type coupling scheme was used, with coupling strengths given by a coupling matrix $\mathbb{C}=(\mathbb{C}_{k,j})$, where $\mathbb{C}_{k,j}$ represents the coupling effect of neuron $j$ to neuron $k$. Phase synchronization was demonstrated to occur at a critical value of $||\mathbb{C}||$ -- in line with the work done by Kuramoto \cite{K1984}. Similar work was done in \cite{GW2019} on a lattice of discrete-Mexican-hat coupled EI-pairs, demonstrating patterns of ordered phases in both space and time that appeared quickly and then evolved slowly.

In some papers, such as \cite{GMW2016} and \cite{GW2019}, the amplitudes of the coupled quasi-cycle oscillators (which correspond to their peak-to-trough voltages) appeared to grow without bound for some parameter regimes in the absence of a bounding function $g$. Although amplitudes of limit cycles were not discussed by Kuramoto in \cite{K1984}, it is important to consider the behaviour of the amplitudes when appraising phase pattern formation in these models. For example, dependence of phase coherence between W-C models of two oscillating neural masses on the ratio of their amplitudes was demonstrated in \cite{DW2011}. And in \cite{GW2019}, amplitude patterns were dissociated from phase patterns at weak coupling strengths but both appeared together with strong coupling. In the latter case, the simulations were run for just long enough to reveal spatial and temporal patterns, but not long enough for amplitudes to grow to extremely large levels. The justification for this procedure was that oscillatory patterns in the brain are seldom stable for very long, usually only for periods of a few hundred milliseconds. Greenwood and Ward \cite{GW2019} suggested that there could be a mechanism that limited oscillatory amplitudes in these cases. In particular they noted that global inhibition or changes in local coupling strengths, among other mechanisms, would promote the development of transient spatial and temporal amplitude and phase patterns among neural oscillations. 

In this paper we demonstrate, via numerical simulations of a Mexican-hat-coupled model similar to that of \cite{GW2019}, that sustained phase patterns occur only when the amplitude processes are unbounded. The introduction of `plastic systemic inhibition,' however, can control the amplitudes while still permitting sustained phase pattern formation to occur. By `plastic systemic inhibition,' we mean a system-dependent regulatory mechanism that targets the intrinsic damping parameter, $\lambda$, of each coupled EI-pair, instead of bounding each voltage by a sigmoid. We also tested a `static' inhibitory mechanism, which is analogous to the plastic mechanism except that the effect on $\lambda$ is fixed instead of adaptive. As it turns out, only with the use of the adaptive plastic systemic inhibitory mechanism do we see bounded amplitudes combined with sustained phase patterns; the static mechanism bounds amplitudes but does not allow for pattern formation in the system. Additionally, we study how the eigenvalues of the total coupled system influence amplitude growth and phase pattern formation, and show that pattern formation is apparently stochastic when using static systemic inhibition.

We conjecture that our plastic systemic inhibitory mechanism may be closely related to homeostatic inhibitory plasticity \cite{VSZCG2011}. Inhibitory plasticity has come into recent interest as a mechanism that can induce both stability and rich dynamics in neural networks. The role of inhibitory plasticity in maintaining the excitation-inhibition balance, the stabilization of recurrent network dynamics, and sensory-response de-correlation is discussed in \cite{S2017}. Synaptic plasticity in inhibitory synapses is explored in \cite{VSZCG2011}, which  explains sparse firing patterns observed in response to natural stimuli, as well as providing a homeostatic mechanism that generates asynchronous and irregular network states. In \cite{HJCL2015}, a local homeostatic inhibitory plasticity scheme is shown to regulate network activity and cause rich and spontaneous dynamics to emerge over a large range of brain configurations, which otherwise have a limited range of dynamic regimes.

In Section \ref{Ito} we summarize the derivation of the phase and amplitude processes with a generic linear-type coupling scheme in the case where every EI-pair has the same parameters. Section \ref{matrix} describes the matrix equations that summarize the dynamics of the coupled system, analogous to \eqref{eimatrix1}. Section \ref{inhibschemes} describes the systemic inhibitory mechanisms. The choice of coupling coefficients, the parameters for the EI-pairs, and the simulation parameters for the numerical results are provided in Section \ref{sysspec}. The results of our simulations are presented in Section \ref{sims}, with a more general discussion in Section \ref{disc}, and the conclusion in Section \ref{concl}.
\section{Model Development}

\subsection{It\^{o} Transformation to Phase and Amplitude Processes}\label{Ito}
We consider a system of $N$ EI-pairs, $(E_1,I_1), ...,(E_k,I_k), ...,(E_N,I_N)$, each of which, in the absence of coupling, obeys \eqref{eimatrix1}, i.e.,
\begin{equation}\label{eimatrix}
d\mathbb{V}_k(t)=[E_k(t),I_k(t)]^\top 
=-\mathbb{A} \mathbb{V}_k(t) dt+\mathbb{N} d\mathbb{W}_k(t),
\end{equation}
where for each $k$, $d\mathbb{W}_k(t)=(dW_k^E,dW_k^I)^\top$, where $\{W_k^E,W_k^I,k=1,...,N\}$ are independent standard Brownian motions. Note that the matrices $\mathbb{A}, \mathbb{N}$ are identical for each EI-pair. We introduce a coupling term, $\mathbb{M}_k (t;\mathbb{C})$, into \eqref{eimatrix}, given by,
\begin{equation}\label{mterm}
\mathbb{M}_k (t;\mathbb{C})=\sum_j \mathbb{C}_{k,j}\mathbb{V}_j(t)
\end{equation}
where $\mathbb{C}=(\mathbb{C}_{k,j})$ is a coupling matrix whose diagonal entries are zero and whose off-diagonal entries, $\{\mathbb{C}_{k,j}\}$, represent the post-synaptic connection of EI-pair $j$ to EI-pair $k$. The introduction of the coupling term modifies \eqref{eimatrix} into,
\begin{equation} \label{system}
d\mathbb{V}_k(t)=\big(-\mathbb{A} \mathbb{V}_k(t) +\mathbb{M}_k (t;\mathbb{C}) \big) dt +\mathbb{N} d\mathbb{W}_k(t),
\end{equation}

Next we transform $-A$ into ``normal form", which is a change of basis that aims to replace the matrix $-A$ in \eqref{system} with a matrix that is written in terms of the intrinsic damping $\lambda$ and frequency $\omega$ of the system \eqref{eimatrix1}. The intrinsic damping and frequency are the real and imaginary parts, respectively, of the matrix $-A$; in other words $-A$ has eigenvalues given by $-\lambda \pm \omega i$. These can be calculated as,
\begin{equation}\begin{split}\label{lambda}
\lambda=-\frac{tr(\mathbb{-A})}{2}=\frac{1-S_{EE}}{2\tau_E}+\frac{1+S_{II}}{2\tau_I},
\end{split}\end{equation}
and,
\begin{equation}\begin{split}\label{omega}
&\omega=\frac{1}{2}\sqrt{tr(-\mathbb{A})^2-4\det (-\mathbb{A})}\\
&=\sqrt{\lambda^2-\frac{S_{IE}S_{EI}+(1-S_{EE})(1+S_{II})}{\tau_E\tau_I}}.
\end{split}\end{equation}
We perform the change of basis $\{\mathbb{Y}_k=\mathbb{Q}_k^{-1}\mathbb{V}_k=(u_k,v_k)^\top \}$ where the matrix $\mathbb{Q}$ is given by,
\begin{equation}\begin{split} \label{matrixq}
\mathbb{Q}=
\begin{pmatrix}
-\omega & \lambda+(-1+S_{EE})\tau_E^{-1} \\
 0& S_{IE}\tau_I^{-1}
\end{pmatrix},
\end{split}\end{equation}
to obtain the matrix $\mathbb{B}$ as,
\begin{equation}\begin{split}\label{bmat}
\mathbb{B}:=\mathbb{Q}^{-1}(-\mathbb{A})\mathbb{Q}=
\begin{pmatrix}
-\lambda &\omega \\
 -\omega &-\lambda
\end{pmatrix}.
\end{split}\end{equation}
The coupling term is stable under the transformation, i.e. $\mathbb{Q}^{-1}\mathbb{M}_k (t;\mathbb{C})\mathbb{Q}=\mathbb{M}_k (t;\mathbb{C})$. Thus the system in the new basis is given by,
\begin{equation}\label{system2}
d\mathbb{Y}_k (t)=(\mathbb{B} \mathbb{Y}_k(t) +\mathbb{M}_k (t;\mathbb{C})) dt+ \mathbb{E} d\mathbb{W}_k(t),
\end{equation}
where $\mathbb{E}=\mathbb{Q}^{-1}\mathbb{N}$.

The derivation of the phase and amplitude processes corresponding to \eqref{system2} are given in Appendix A. In the case where every EI-pair has identical parameters, these processes are given respectively by,
\begin{widetext}
\begin{equation}\label{zeq}
dZ_k=\bigg( \frac{a^2+b^2+c^2-(a^2+b^2) \cos (\theta_k(t))^2}{2Z_k}
\frac{-c^2 \sin (\theta_k(t))^2-bc\sin(2\theta_k(t))}{2Z_k}
-\lambda Z_k +\sum_j \mathbb{C}_{k,j}Z_j\cos(\theta_k-\theta_j) \bigg) dt+dR_k.
\end{equation}
and,
\begin{equation}\label{thetaeq}
d\theta_k=\bigg(bc\frac{1-2\cos(\theta_k(t))^2}{Z_k^2}-\omega +\sum_j \mathbb{C}_{k,j} \frac{Z_j}{Z_k}\sin(\theta_j-\theta_k) \bigg) dt+dS_k,
\end{equation}
\end{widetext}
where,
\begin{equation}\label{acoef}
a=\frac{-\sigma_E}{\omega\tau_E}
\end{equation}
\begin{equation}\label{ccoef}
c=\frac{\sigma_I}{S_{IE}},
\end{equation}
\begin{equation}\label{bcoef}
b=\frac{-1+S_{EE}+\lambda\tau_E}{\omega \tau_E}c,
\end{equation}
and where the noise terms are given by,
\begin{align}\begin{split}\label{rnoise}
&dR_k=a\cos(\theta_k(t)) dW_k^E \\
&+ (b\cos(\theta_k(t))+c \sin (\theta_k(t)))dW_k^I,
\end{split}\end{align}
and,
\begin{align}
\begin{split}\label{snoise}
&dS_k(t)=\\
&\frac{-a\sin(\theta_k(t))dW_k^E+(c\cos(\theta_k(t))
-b\sin(\theta_k(t)))dW_k^I}{Z_k}.
\end{split}\end{align}

\subsection{Matrix Form}\label{matrix}
We write the system of $N$ coupled EI-pairs as a matrix equation. We let $\mathcal{Y}=(\mathbb{Y}_1^\top,...,\mathbb{Y}_N^\top)^\top$ and $\mathcal{W}=(\mathbb{W}_1^\top,...,\mathbb{W}_N^\top)^\top$. The block matrix $\mathcal{D}$, given by,
\begin{equation}\begin{split} \label{dmatrix}
\mathcal{D}=
\begin{pmatrix}
\mathbb{B} & 0 & \hdots & 0& 0 \\
0 & \mathbb{B} & 0 & \hdots & 0 \\
\vdots & 0 &\ddots&& \vdots\\
0&\vdots&&\mathbb{B}& 0\\
0 &0  &\hdots&0&\mathbb{B} 
\end{pmatrix},
\end{split}\end{equation}
captures the individual uncoupled deterministic dynamics of each EI-pair as in \eqref{eimatrix} (without the noise term). Next, the block matrix $\mathcal{K}(\mathbb{C})$, given by,
\begin{equation}\begin{split} \label{kmatrix}\resizebox{0.88\columnwidth}{!}{$
\mathcal{K}(\mathbb{C})=
\begin{pmatrix}
0 & \mathbb{C}_{1,2}\mathbb{I}_2 & \mathbb{C}_{1,3}\mathbb{I}_2 & \hdots & \mathbb{C}_{1,N}\mathbb{I}_2 \\
\mathbb{C}_{2,1}\mathbb{I}_2 & 0 & \mathbb{C}_{2,3}\mathbb{I}_2 & \hdots & \mathbb{C}_{2,N}\mathbb{I}_2 \\
\mathbb{C}_{3,1}\mathbb{I}_2 & \mathbb{C}_{3,2}\mathbb{I}_2 &0& \hdots & \mathbb{C}_{3,N}\mathbb{I}_2 \\
\vdots & \vdots &\vdots & \ddots & \vdots\\
\mathbb{C}_{N,1}\mathbb{I}_2& \mathbb{C}_{N,2}\mathbb{I}_2 &\mathbb{C}_{N,3}\mathbb{I}_2& \hdots & 0 
\end{pmatrix},$}
\end{split}\end{equation}
captures the coupling aspect of the system in \eqref{system}, where $\mathbb{I}_2$ is the $2 \times 2$ identity matrix. The matrix $\mathcal{E}$, given by,
\begin{equation}\begin{split}
\mathcal{E}=
\begin{pmatrix}
\mathbb{E} & 0 & \hdots & 0& 0 \\
0 & \mathbb{E} & 0 & \hdots & 0 \\
\vdots & 0 &\ddots&& \vdots\\
0&\vdots&&\mathbb{E}& 0\\
0 &0  &\hdots&0&\mathbb{E} 
\end{pmatrix},
\end{split}\end{equation}
provides the correct transformation of the i.i.d. Brownian motions in $\mathcal{W}$. Thus the system of $N$ EI-pairs governed by \eqref{system} can be written as the matrix equation,
\begin{equation}\begin{split} \label{transmat}\resizebox{0.88\columnwidth}{!}{$
d\mathcal{Y}(t)=\big(\mathcal{D}\mathcal{Y}(t)+\mathcal{K}(\mathbb{C})\mathcal{Y}(t) \big) dt+\mathcal{E} d \mathcal{W}(t).$}
\end{split}\end{equation}
Note that the matrix $\mathcal{D}$ is a constant matrix, defined by the choice of parameters in \eqref{eimatrix}, whereas the matrix $\mathcal{K}(\mathbb{C})$ depends both on the parameters in \eqref{eimatrix}, and on the choice of coupling matrix $\mathbb{C}$. In our numerical simulations and analyses that follow, we are concerned with the deterministic temporal eigenvalues of our coupled system, which are the eigenvalues of the block matrix $\mathcal{L}(\mathbb{C})$, given by,
\begin{align*}\label{systemmat}
\mathcal{L}(\mathbb{C})=\mathcal{D}+\mathcal{K}(\mathbb{C})
\end{align*}
\begin{align}
=\begin{pmatrix}
\mathbb{B} & \mathbb{C}_{1,2}\mathbb{I}_2 & \mathbb{C}_{1,3}\mathbb{I}_2 & \hdots & \mathbb{C}_{1,N}\mathbb{I}_2 \\
\mathbb{C}_{2,1}\mathbb{I}_2 & \mathbb{B} & \mathbb{C}_{2,3}\mathbb{I}_2 & \hdots & \mathbb{C}_{2,N}\mathbb{I}_2 \\
\mathbb{C}_{3,1}\mathbb{I}_2 & \mathbb{C}_{3,2}\mathbb{I}_2 &\mathbb{B}& \hdots & \mathbb{C}_{3,N}\mathbb{I}_2 \\
\vdots & \vdots &\vdots & \ddots & \vdots\\
\mathbb{C}_{N,1}\mathbb{I}_2& \mathbb{C}_{N,2}\mathbb{I}_2 &\mathbb{C}_{N,3}\mathbb{I}_2& \hdots & \mathbb{B} 
\end{pmatrix},
\end{align}
\subsection{Systemic Inhibitory Schemes}\label{inhibschemes}
Now we introduce the systemic inhibitory mechanisms investigated in this paper. The first is a `static' systemic inhibitory mechanism where we subtract $\delta \mathbb{I}_{2N \times 2N}$, for a systemic inhibition parameter $\delta>0$, from \eqref{transmat} leading to the matrix equation for the statically inhibited system given by,
\begin{equation}\label{staticinhib}
d\mathcal{Y}(t)=\bigg(\big(\mathcal{L}(\mathbb{C})-\delta \mathbb{I}_{2N \times 2N}\big)\mathcal{Y}(t)\bigg)dt +\mathcal{E}d\mathcal{W}(t).
\end{equation}

We also explore two `plastic' systemic inhibitory mechanisms, where we instead subtract $\delta \xi(t)$, for a systemic inhibition parameter $\delta>0$, from \eqref{transmat} resulting in the matrix equation,
\begin{equation}\label{plasticinhib}
d\mathcal{Y}(t)=\bigg(\big(\mathcal{L}(\mathbb{C})-\delta \xi(t) \big)\mathcal{Y}(t)\bigg)dt +\mathcal{E}d\mathcal{W}(t).
\end{equation}
The first plastic systemic inhibition mechanism is what we will call `binary-type' plastic systemic inhibition, where $\xi$ is given by,
\begin{align}\label{xi1}
\xi(t)=
\begin{pmatrix}
T_1 (t)\mathbb{I}_2 & 0 & \hdots & 0& 0 \\
0 & T_2 (t)\mathbb{I}_2 & 0 & \hdots & 0 \\
\vdots & 0 &\ddots&& \vdots\\
0&\vdots&&T_{N-1} (t)\mathbb{I}_2& 0\\
0 &0  &\hdots&0&T_N (t)\mathbb{I}_2
\end{pmatrix},
\end{align}
where $\mathbb{I}_2$ is the $2 \times 2$ identity matrix and where,
\begin{align*}
T_i (t)=1\{Z_i(t)>z^*\}=
\begin{cases}
1 &\text{if}\,\, Z_i(t)>z^*\\
0 & \text{otherwise}\\
\end{cases},
\end{align*}
for a plastic amplitude threshold $z^*$. The second plastic mechanism we explored is what we will call `saturation-type' plastic systemic inhibition, where $\xi(t)$ is given by,
\begin{align}\label{xi2}
\xi(t)=
\begin{pmatrix}
U_1 (t) \mathbb{I}_2& 0 & \hdots & 0& 0 \\
0 & U_2 (t)\mathbb{I}_2& 0 & \hdots & 0 \\
\vdots & 0 &\ddots&& \vdots\\
0&\vdots&&U_{N-1} (t)\mathbb{I}_2& 0\\
0 &0  &\hdots&0&U_N (t)\mathbb{I}_2
\end{pmatrix},
\end{align}
where
\begin{align*}\label{satmech}
U_i (t)=\frac{1}{1+\max\{0,z^*-Z_i(t)\}}.
\end{align*}
As we will see in Section \ref{sims}, both plastic inhibitory mechanisms have the desired effect of keeping amplitudes stochastically bounded while allowing pattern formation to occur in the phases. The saturation-type inhibition mechanism is more biologically plausible, however, as the amplitude threshold $z^*$ in that mechanism corresponds to a maximum, or `saturated,' rate of firing of a neuron or of a neural population. 

When choosing a systemic inhibition parameter $\delta$ so that the resulting system has a specific maximum eigenvalue real-part when systemic inhibition is applied, it is practical to instead first pick a target maximal eigenvalue real-part for the matrix $\mathcal{L}(\mathbb{C})-\delta \mathbb{I}_{2N \times 2N}$ in \eqref{staticinhib}, or $\mathcal{L}(\mathbb{C})-\delta \xi (t)$ in \eqref{plasticinhib}, and then choose $\delta$ accordingly. We did this because the eigenvalue real parts provide a heuristic as to whether the system remains close to or deviates far from the origin ($Z=0$). For example, one would expect that if the eigenvalues of the matrix $\mathcal{L}(\mathbb{C})-\delta \mathbb{I}_{2N \times 2N}$ all have negative real part, then the system will remain bounded -- which is exactly what we are trying to achieve by raising the systemic inhibition when the amplitude is too large. Conversely, we expect amplitudes to be unbounded when there is at least one eigenvalue with a positive real part. In the case where $\mathcal{L}(\mathbb{C})$ is diagonalizable, subtracting $\delta \mathbb{I}_{2N \times 2N}$ will subtract $\delta$ from all of the eigenvalues. Hence when $\mathcal{L}(\mathbb{C})$ is diagonalizable, we can numerically find the maximal eigenvalue real part, say $\lambda^*$, and pick a target maximal eigenvalue real part, say $\Lambda^*$, and set $\delta=\lambda^*-\Lambda^*$. Then $\mathcal{L}(\mathbb{C})-\delta \mathbb{I}_{2N \times 2N}$ will have maximal eigenvalue real-part $\Lambda^*$. It is not as clear whether this method would work in general when the matrix $\mathcal{L}(\mathbb{C})$ is not diagonalizable, but in our simulations that follow we re-compute the maximal eigenvalue real part to ensure this works as intended. We will hereafter refer to $\Lambda^*$ as the `adjusted eigenvalue bound.' See Section \ref{disc} for a discussion on the biological interpretation of these systemic inhibitory mechanisms.

\subsection{Parameter Specifications and Mexican Hat Coupling}\label{sysspec}
We consider our EI-pairs regarded as a one dimensional ring-lattice with uniform spacing. That is, EI-pair $k$ has the integer position $k$ on the ring-lattice $\{1,...,N\} \mod N$. Every EI-pair is set to have parameters $\tau_E=0.003,\tau_I=0.006,S_{EE}=1.5,S_{II}=0.1,S_{IE}=4,$ and $S_{EI}=1$, whereas the damping rates (real parts of eigenvalues of $\mathbb{A}$) and intrinsic frequencies (imaginary parts of eigenvalues of $\mathbb{A}$) were adjusted to fit the other parameters according to \eqref{lambda} and \eqref{omega}. Population size is $N=100$. Although the specifications of our simulations closely follow that of \cite{GW2019}, that work did not specify the constants $\sigma_E$ and $\sigma_I$, and so we used $\sigma_E=\sigma_I=12$ as in \cite{GMW2016}. As mentioned in \cite{GMW2016}, these sets of parameters yield a narrow-band quasi-cycle oscillation at approximately 70 Hz, which belongs to the gamma frequency range of 30-80 Hz which has been shown to be important in neural oscillations \cite{GMW2015}. The Euler-Maruyama method was used to numerically solve the stochastic differential equations in \eqref{zeq} and \eqref{thetaeq} with time step $ \Delta t=5 \cdot 10^{-5}$

In this paper we use the `Mexican hat' function, which is a difference of Gaussian density functions, as a kernel in our coupling scheme. It is given by,
\begin{equation}\begin{split} \label{mexhat}
&m(x)=b_1 \exp \big[-(\frac{x}{d_1})\big]-b_2 \exp \big[-(\frac{x}{d_2}) \big],\\
& b_2>b_1, \hspace{0.1cm} d_2>d_1,
\end{split}\end{equation}
where $b_1$ and $b_2$ are the amplitudes and $d_1$ and $d_2$ the diffusive parameters for two Gaussians. That is, we define the coupling matrix $\mathbb{C}$ by,
\begin{equation}\begin{split}\label{cscheme}
\mathbb{C}_{k,j}=Cm(k-j),
\end{split}\end{equation}
where $C>0$ is an adjustable parameter. Note that the difference $k-j$ in $m(k-j)$ is actually the minimum distance between points $k$ and $j$ respecting the ring structure of the lattice. The coupling term, $\mathbb{M}_k(t;\mathbb{C})$, in \eqref{system} is thus given by,
\begin{equation}\begin{split}
\mathbb{M}_k(t;\mathbb{C})=\sum_{j=1}^N C m(k-j)\mathbb{V}_j(t).
\end{split}\end{equation}
The spacing of the EI-pairs can be defined implicitly by a scaling of the parameters $d_1$ and $d_2$ in \eqref{mexhat}, and for this reason we let the EI-pairs be spaced one unit apart in \eqref{cscheme}. In \cite{GW2019}, $b_2$ and $d_1$ were fixed to be 1 (in order to make the analysis more tractable), and then $b_1=1.3$ and $d_2=1.5$ were determined to be suitable parameters for achieving pattern formation. Unlike in \cite{GW2019}, we employ unit distances between EI-pairs instead of a separation distance of 0.2 per EI-pair, and consider diffusion parameters roughly five times greater in order to account for this difference. In \cite{GW2019}, patterns were found to form with $C=1$ and more so at $C=8$, and so here we chose $C=8$.


In order to measure the degree of spatial pattern formation in the simulated phases, we plot the spatial sample entropy as a function of time. The version of sample entropy that we use in this paper is defined in \cite{RM2000} -- which provides background motivation as well as technical details. This sample entropy is an unbiased estimator of the `R\'{e}nyi entropy of order 2.'  Details are given in the `Sample Entropy' section in the Supplementary Material. In our context, high values of entropy suggest absence of a pattern, whereas low values of entropy suggest the presence of a pattern.  We computed sample entropies using an algorithm given in \cite{RBL2009} via an implementation provided by \cite{samplealgorithm} (wherein we set tolerance parameter $r=1$, embedding dimension parameter $m=1$, and use Chebychev distance measure). Sample entropy provides a useful heuristic to aid visual inspection and to quantify the degree of pattern formation.  



\begin{figure}[ht!]
\begin{center}
\includegraphics[width=3.6 in]{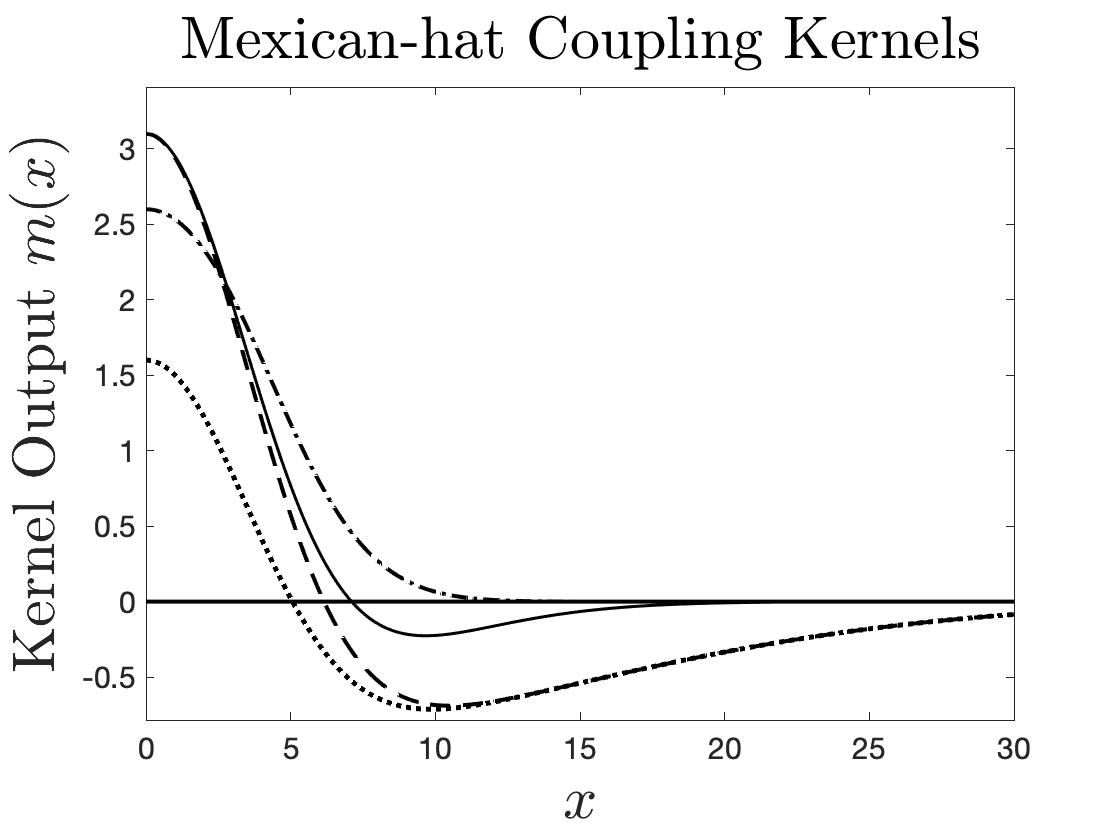}
\end{center}
\caption{Mexican hat coupling kernels, as in \eqref{mexhat}, with couplers: (A) $(b_1,d_2)=(2.6,19.1)$ with max eigenvalue real-part 115.8 (dotted curve), (B) $(b_1,d_2)=(4.1,9.1)$ with max eigenvalue real-part 137.8 (solid curve), (C) $(b_1,d_2)=(3.6,3.6)$ with max eigenvalue real-part 175.1 (dotted and dashed curve), (D) $(b_1,d_2)=(4.1,19.1)$ with max eigenvalue real-part 189 (dashed curve). The parameters $b_2$ and $d_1$ are set to 1 and 5, respectively, for each coupler. The line $y=0$ is dashed for reference. It is relevant to note that the more realistic couplers in this figure are coupler A (dotted curve) and D (dashed curve) as both have a balance of excitation and inhibition. By contrast, coupler B (solid curve) has very little inhibition compared to its excitation and coupler C (dotted and dashed curve) has almost no inhibition.}
\label{fig:mexhat1}
\end{figure}
\begin{figure}[ht!]
\begin{center}
\includegraphics[width=3.6 in]{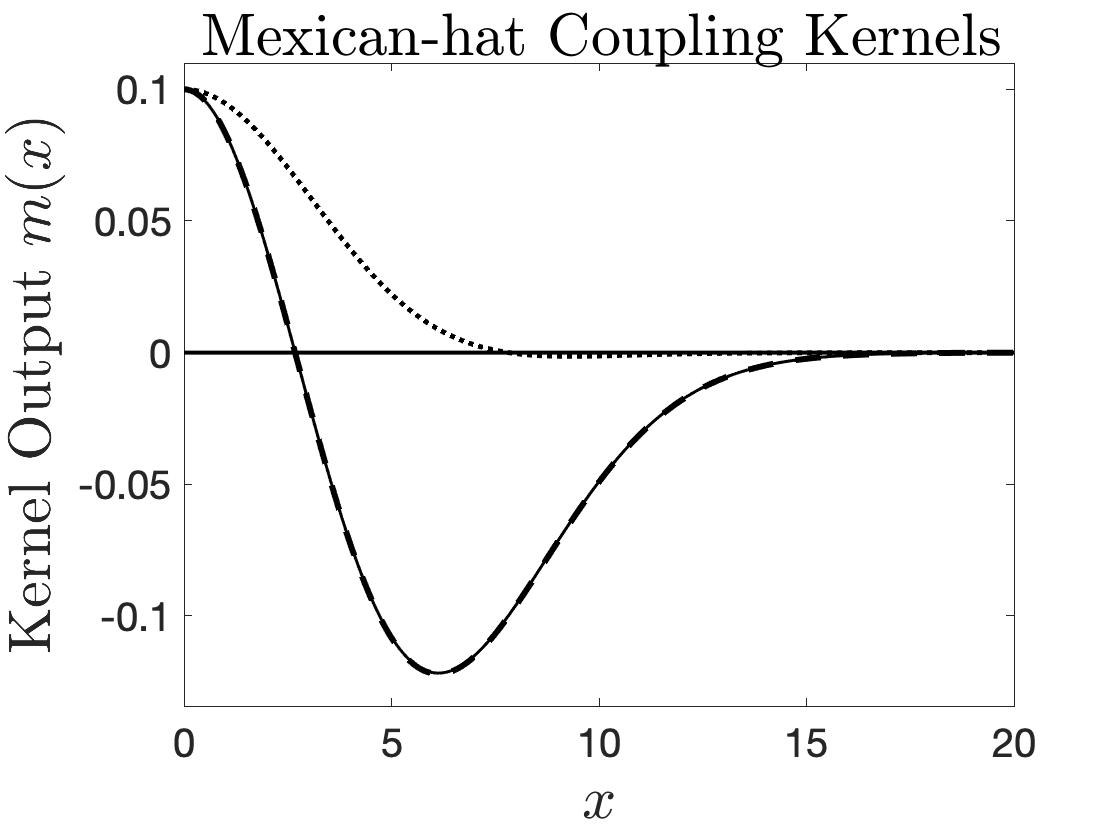}
\end{center}
\caption{Mexican hat coupling kernels, as in \eqref{mexhat}, with couplers: (E) $(b_1,d_2)=(1.1,5.1)$ with max eigenvalue real-part -3.461 (dotted curve), (F) $(b_1,d_2)=(1.1,6.1175)$ with max eigenvalue real-part -0.04708 (solid curve), (G) $(b_1,d_2)=(1.1,6.12)$ with max eigenvalue real-part 0.01516 (dashed curve). The parameters $b_2$ and $d_1$ are set to 1 and 5, respectively, for each coupler. The line $y=0$ is dashed for reference. It is relevant to note that coupler E (dotted curve) is unrealistic as it has essentially no inhibition, whereas couplers F (solid curve) and G (dashed curve) both have some balance of excitation and inhibition.} 
\label{fig:mexhat2}
\end{figure}

For the simulations that follow in this paper we consider Mexican hat coupling kernels for 7 sets of parameters $(b_1,b_2,d_1,d_2)$, but with the values of $b_2=1, d_1=5$ held constant throughout. We enumerate the subsequent couplers (i.e. the matrix $\mathbb{C}$ for each system) as couplers A through G, as displayed in Table 1. The parameters control the shape of the Mexican Hat function (see Figs. \ref{fig:mexhat1} and \ref{fig:mexhat2}) and thus the visual appearance of the phase patterns; larger values of $d_2$ relative to $d_1$ imply a wider reach of the Mexican hat, and different ratios $b_1/b_2$ determine the relative amount of inhibition around the  excitatory centre. Our focus here is on the formation of phase patterns rather than on their visual appearance, so we won't comment further on the character of the phase patterns although their differences will be apparent and can be related to the parameters of the respective Mexican hats.

The purpose of choosing a diverse range of couplers was to investigate whether results regarding the presence or absence of pattern formation vary for different values of coupler parameters that had a variety of max eigenvalue real-parts. Some couplers in Table 1 are more realistic than others; e.g. coupler A has a balance of excitation and inhibition, whereas coupler C has almost no inhibition. Nevertheless, we were interested in testing for pattern formation across a variety of dynamic regimes, and so we also simulated the more unrealistic couplers. As it turns out the differences in the absence or presence of pattern formation were limited. Couplers A through D behaved essentially the same with regard to pattern formation, but differently than couplers E through G -- the three of which behaved essentially the same. In Section 3 we present only the simulations for couplers A and F, relegating the others to the Supplementary Material.

%

\begin{table}
\begin{center}\resizebox{\columnwidth}{!}{
\begin{tabular}{|c||c|c|c|}                                                                                                       
\hline                                                                                                                               
Pair&$b_1$&$d_2$&Max Eigenvalue Real-part \\
\hline
A&2.6&19.1&115.8\\
B&4.1&9.1&137.8\\
C&3.6&3.6&175.1\\
D&4.1&19.1&189.0\\
E&1.1&5.1&-3.462\\
F&1.1&6.1175&-0.0004060\\
G&1.1&6.12&0.01516\\
\hline                                                                                                                                    
\end{tabular}    }
\caption{Parameter values in the coupling kernel \eqref{mexhat} for each of the 7 couplers.}
\label{params1}
\end{center}
\end{table}

\section{Creating Sustained Phase Patterns with Bounded Amplitudes}\label{sims}
We simulated \eqref{zeq} and \eqref{thetaeq} with the Mexican hat coupling parameters indicated in Table \ref{params1} for a range of time periods depending on the parameters and the results of the simulations. Time periods were generally shorter when amplitudes grew without bound, and longer when amplitudes remained bounded, in order to demonstrate the outcome convincingly. In what follows we display only a subset of the simulation plots that illustrate our overall results. Results not displayed but described can be found in the Supplementary Material.

\subsection{Investigating Phase Patterns without Systemic Inhibition}
\begin{figure*}
\begin{center}
\resizebox{2\columnwidth}{!}{\includegraphics{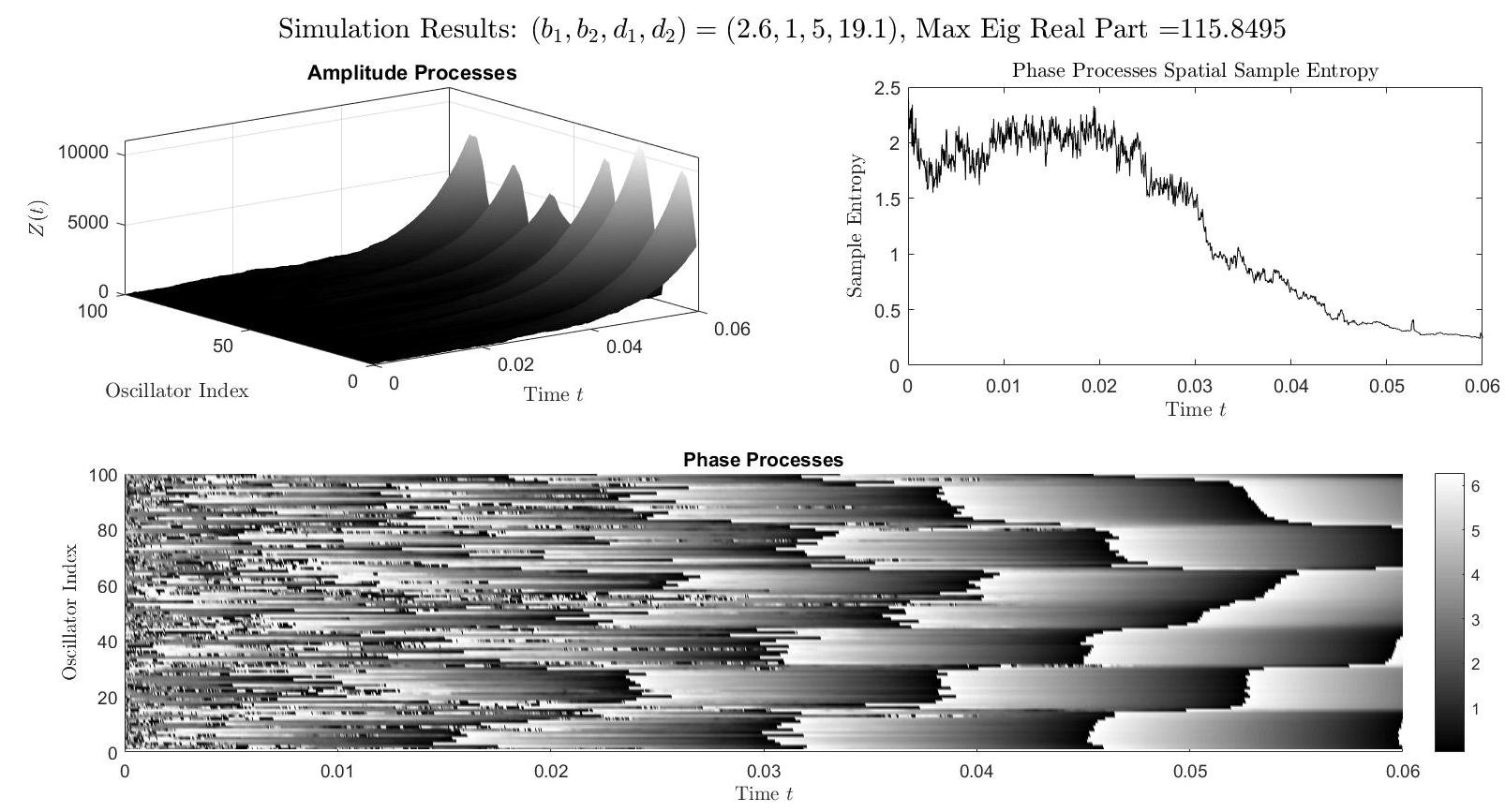}}
\end{center}
\caption{Phase and amplitude process simulation with Mexican Hat coupling \eqref{mexhat} with coupler A; $(b_1,b_2,d_1,d_2)=(2.6,1,5,19.1)$. The maximum eigenvalue real part of the matrix $\mathcal{L}(\mathbb{C}$), given by \eqref{systemmat}, with these parameters was 115.8. Initial amplitudes were selected from a uniform distribution on $(0,1]$ and initial phases from a uniform distribution on $[0,2\pi]$.}
\label{fig:sim1}
\end{figure*}

Couplers A through D all induce positive maximum eigenvalue real-parts with an order of magnitude of $10^2$ for their respective matrices $\mathcal{L}(\mathbb{C})$ \eqref{systemmat}. The initial conditions for simulations without systemic inhibition using these pairs have amplitudes taken from a uniform distribution on $(0,1]$, and phases taken from a uniform distribution on $[0,2\pi]$. We ran these simulations in order to illustrate clearly how the model behaves when maximum eigenvalue real-parts are large and positive. Under these conditions amplitudes grow very quickly and apparently without bound. A representative example of these results is displayed in Fig. \ref{fig:sim1} for coupler A (similarly for couplers B through D; see Figs. S1-S3). A clear pattern develops over a short time in the phases as well as in the amplitudes. Sample entropy of the phases is high initially but quickly decreases to a low level as the phase pattern develops and evolves. In all of these four simulations, the amplitudes grow to extreme orders of magnitude in a short period of time. As expected, the systems with larger maximal eigenvalue real-parts tend to have amplitudes with greater orders of magnitude when compared on equal time scales; e.g. the system with coupler A has maximal eigenvalue real part 115.8 and amplitudes that reach a magnitude of $10^4$ at $t=0.06$ (Fig. \ref{fig:sim1}) whereas the system with coupler D has maximal eigenvalue real part of 189.0 and amplitudes that reach magnitudes of $10^5$ at $t=0.06$ (Fig. S3).

\begin{figure*}
\begin{center}
\resizebox{2\columnwidth}{!}{\includegraphics{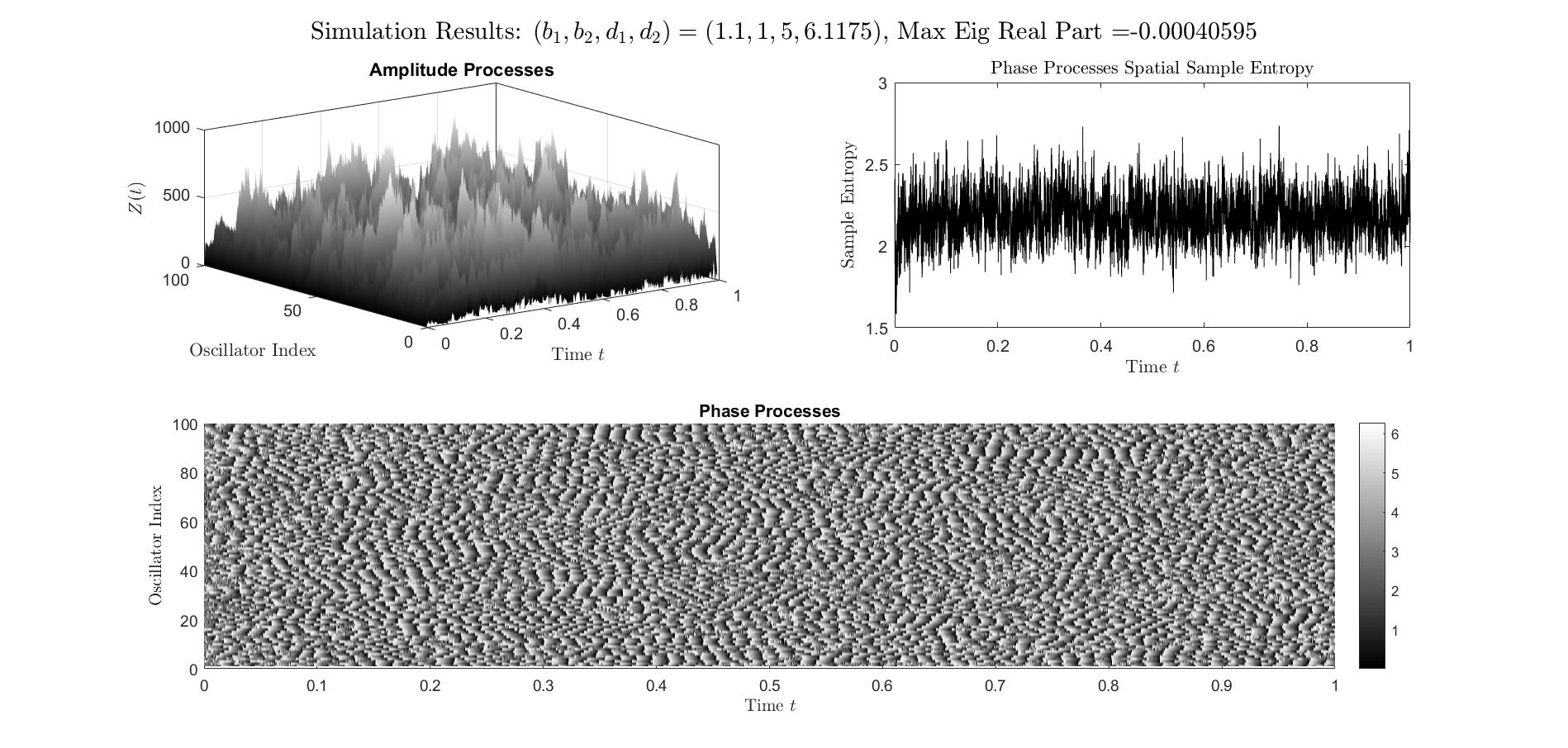}}
\end{center}
\caption{Phase and amplitude process simulation with Mexican Hat coupling \eqref{mexhat} with coupler F; $(b_1,b_2,d_1,d_2)=(1.1,1,5,6.1175)$. The maximum eigenvalue real part of the matrix $\mathcal{L}(\mathbb{C}$), given by \eqref{systemmat}, with these parameters was -0.00040595. Initial amplitudes were selected from a uniform distribution on $(0,1]$ and initial phases from a uniform distribution on $[0,2\pi]$.}
\label{fig:sim6}
\end{figure*}

We also simulated \eqref{zeq} and \eqref{thetaeq} for couplers E through G without systemic inhibition. Fig. \ref{fig:sim6} displays the results using coupler F (similar results for E and G in Figs. S4-S5, respectively), which has a very small negative real part of the eigenvalues, and thus some damping. In this case the amplitudes do not grow without bound but remain in the vicinity of $2\cdot10^2$, even though we ran the simulation for 1 sec.  Unlike the results for large positive real eigenvalues, no clear pattern develops in the phases or in the amplitudes, and the sample entropy remains high. Even when the real part of the eigenvalues is positive, but very small, this result is the same (Fig. S5).

\begin{figure*}
\begin{center}
\resizebox{2\columnwidth}{!}{\includegraphics{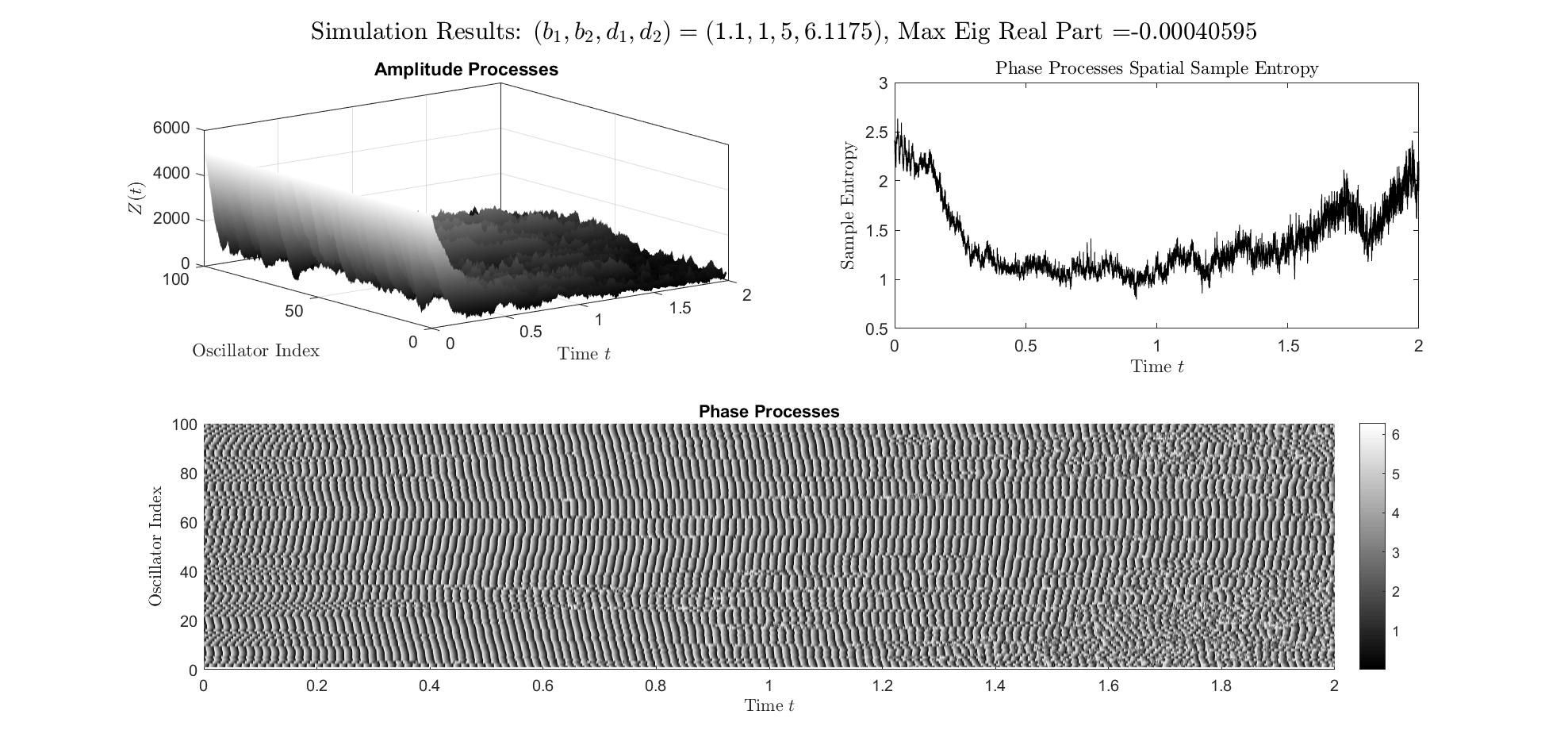}}
\end{center}
\caption{Phase and amplitude process simulation with Mexican Hat coupling \eqref{mexhat} with coupler F; $(b_1,b_2,d_1,d_2)=(1.1,1,5,6.1175)$. The sample entropy of the spatial phase distribution is also plotted for every time point with tolerance $\tau=1$ and embedding dimension $m=1$. The maximum eigenvalue real part of the matrix $\mathcal{L}(\mathbb{C}$), given by \eqref{systemmat}, with these parameters is -0.00040595. Initial amplitudes are selected from a uniform distribution on $(0,1]+5000$ and initial phases from a uniform distribution on $[0,2\pi]$.}
\label{fig:sim9}
\end{figure*}

Additional simulations were run for couplers E through G with the same parameters except that the initial amplitudes were sampled from a uniform distribution on (0,1]+5000, and the simulations were run for a longer time period: 2-3 sec. Figure \ref{fig:sim9} displays the results for coupler F (similar results for couplers E and G in Figs. S6-S7, respectively). When the amplitudes are large, at the beginning of the run, clear spatial patterns similar to those observed in Fig. \ref{fig:sim1} (and Figs. S1-S3) develop in the phases. These patterns decay over time, however, as the amplitudes damp to near zero. The sample entropies of the phases decrease initially as the spatial pattern appears but then increases again as the pattern decays. 

The results of the simulations illustrated in Fig. \ref{fig:sim9} (and in Figs. S6-S7) imply that the lack of pattern formation observed in Fig. \ref{fig:sim6} (and in Figs. S4-S5) is not because the simulations were run for an insufficient time period for large amplitudes to develop. We have seen that phase pattern formation occurs when the amplitudes appear to increase without bound. Therefore, if patterns were to form at some point in time, we would also expect large amplitudes to develop at or before that time point. But when we set the amplitudes to large initial values, but with small positive or negative real parts of the eigenvalues, the amplitudes rapidly decrease from those large initial values, and the phase patterns that are initially present disappear. It thus seems that the appearance of phase patterns in this model is dependent on the sustaining of large amplitudes.

\subsection{Static Systemic Inhibition Prevents and Destroys Phase Patterns}

\begin{figure*}
\begin{center}
\resizebox{2\columnwidth}{!}{\includegraphics{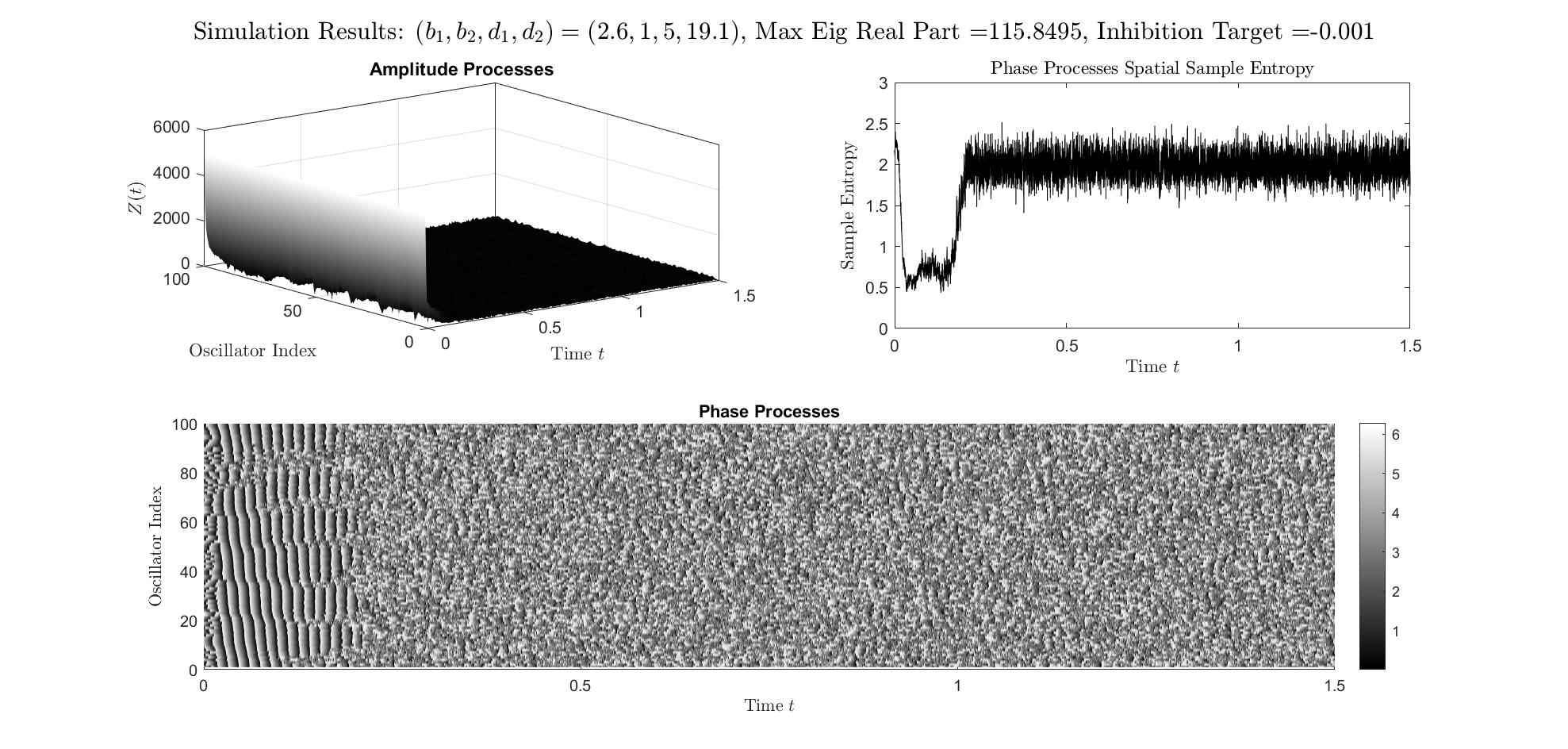}}
\end{center}
\caption{Phase and amplitude process simulation with Mexican Hat coupling \eqref{mexhat} with coupler A; $(b_1,b_2,d_1,d_2)=(2.6,1,5,19.1)$. The maximum eigenvalue real part of the matrix $\mathcal{L}(\mathbb{C}$), given by \eqref{systemmat}, with these parameters is 115.8. Initial amplitudes are selected from a uniform distribution on $(0,1]+5000$ and initial phases from a uniform distribution on $[0,2\pi]$. Static systemic inhibition \eqref{staticinhib} was used with $\delta=115.8505$ corresponding to an adjusted eigenvalue bound of $\Lambda^*=-10^{-3}$.}
\label{fig:sim11}
\end{figure*}

To study the effect of static systemic inhibition to bound amplitudes, we simulated the model for couplers A through D, as specified in Table \ref{params1}, with initial amplitudes taken from a uniform distribution on $(0,1]$, and phases taken from a uniform distribution on $[0,2\pi]$. The adjusted eigenvalue bound $\Lambda^*=-10^{-3}$ (defined in Section \ref{sysspec}) was used for each simulation. This bound was chosen so that the maximal eigenvalue real parts would all be negative -- ensuring that the amplitudes remain bounded -- but close to zero to minimize the magnitude of the inhibitory effect (i.e. to minimize $\delta$ which is defined in Section \ref{sysspec}). Fig. \ref{fig:sim11} displays a representative example of these results for coupler A (similar results for couplers B through D; see Figs. S8-S10). The results are qualitatively similar to Fig. \ref{fig:sim9} (and to Figs. S6-S7), however the phase patterns seem to disappear faster as the amplitude damps for these simulations with static systemic inhibition.

\begin{figure*}
\begin{center}
\resizebox{2\columnwidth}{!}{\includegraphics{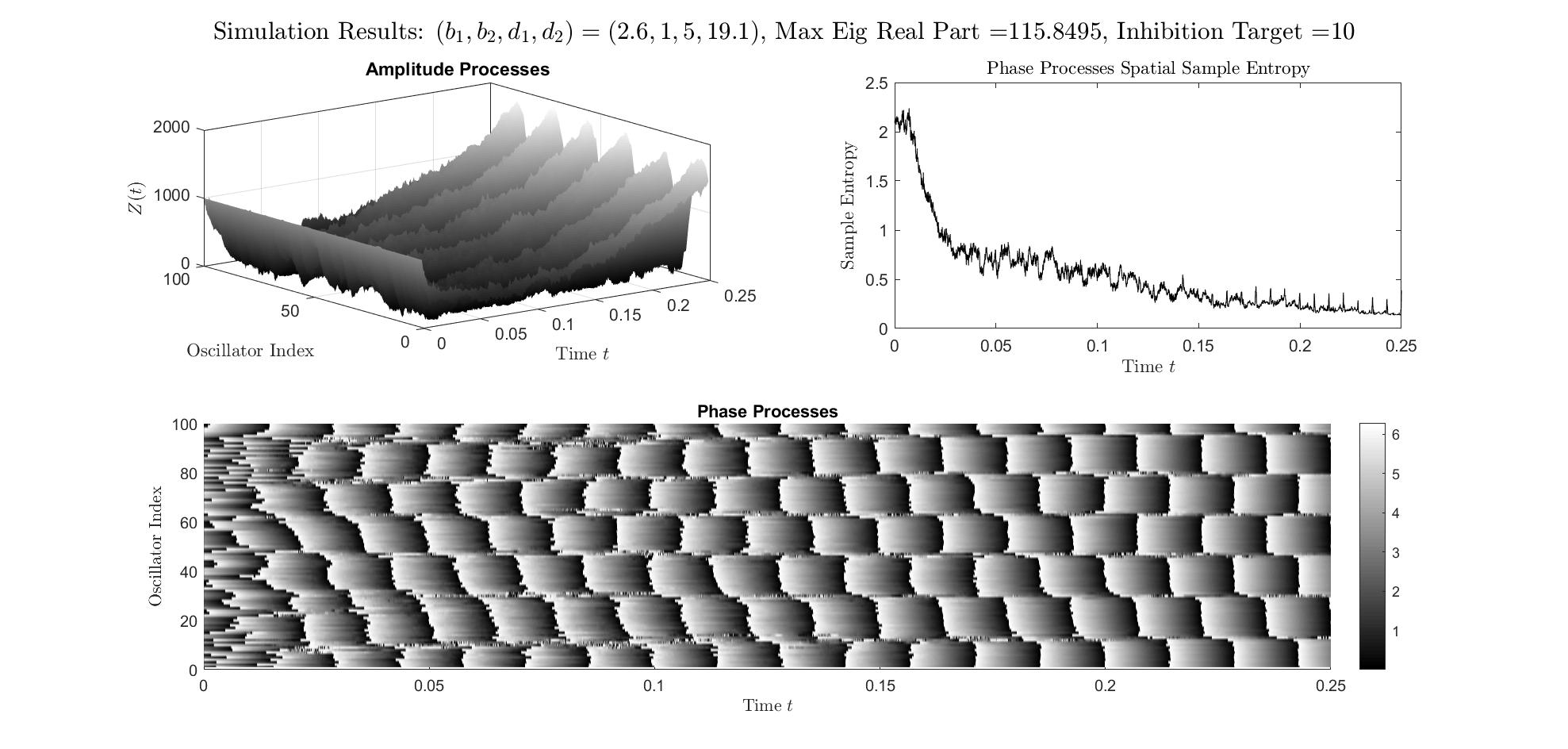}}
\end{center}
\caption{Phase and amplitude process simulation with Mexican Hat coupling \eqref{mexhat} with coupler A; $(b_1,b_2,d_1,d_2)=(2.6,1,5,19.1)$. The maximum eigenvalue real part of the matrix $\mathcal{L}(\mathbb{C}$), given by \eqref{systemmat}, with these parameters is 115.8. Initial amplitudes are selected from a uniform distribution on $(0,1]+1000$ and initial phases from a uniform distribution on $[0,2\pi]$. The systemic inhibition parameter $\delta=105.9$ corresponding to an adjusted eigenvalue bound of $\Lambda^*=10$.}
\label{fig:sim15}
\end{figure*}

\begin{figure*}
\begin{center}
\resizebox{2\columnwidth}{!}{\includegraphics{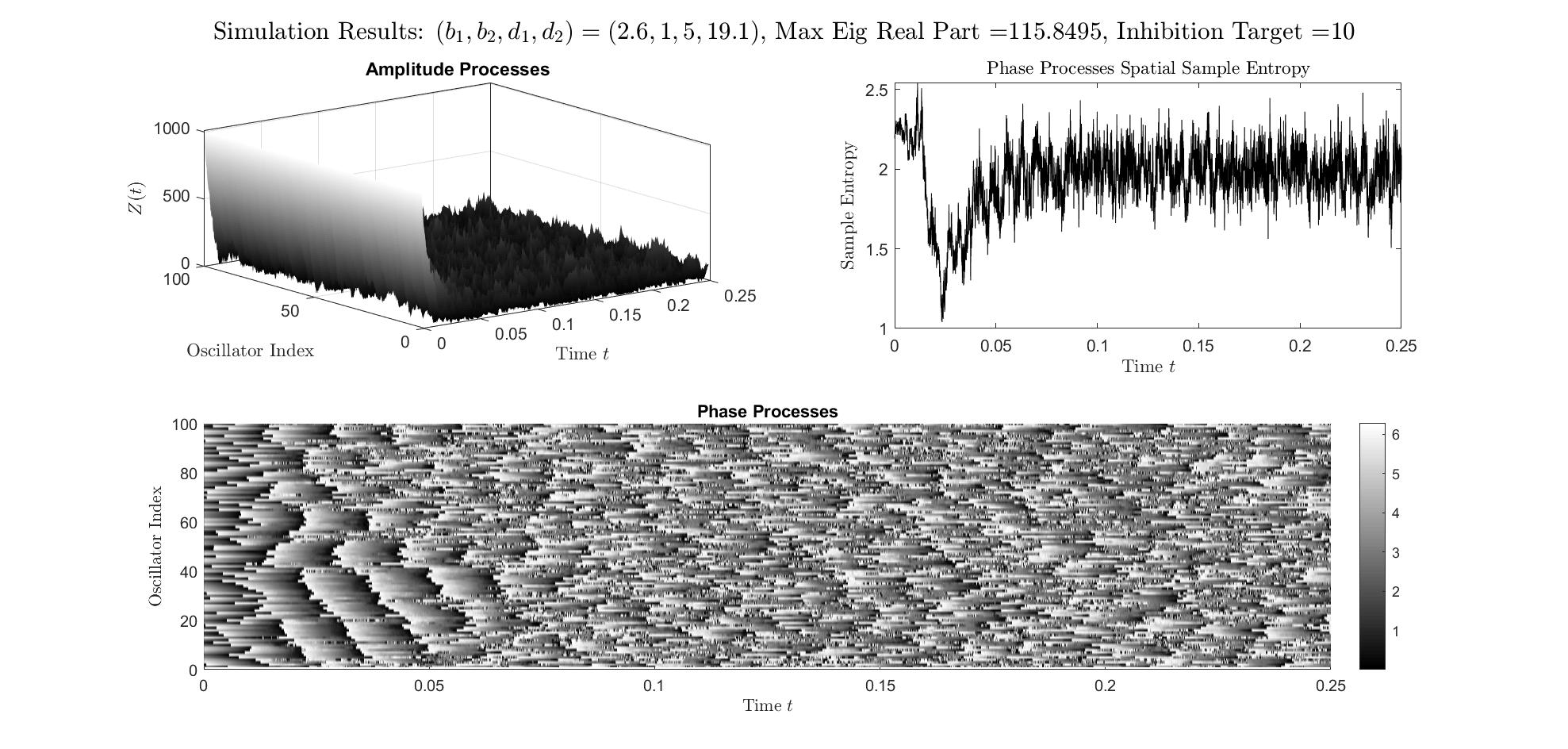}}
\end{center}
\caption{Phase and amplitude process simulation with Mexican Hat coupling \eqref{mexhat} with coupler A; $(b_1,b_2,d_1,d_2)=(2.6,1,5,19.1)$. The maximum eigenvalue real part of the matrix $\mathcal{L}(\mathbb{C}$), given by \eqref{systemmat}, with these parameters is 115.8. Initial amplitudes are selected from a uniform distribution on $(0,1]+1000$ and initial phases from a uniform distribution on $[0,2\pi]$.  Static systemic inhibition \eqref{staticinhib} was used with $\delta=105.9$ corresponding to an adjusted eigenvalue bound of $\Lambda^*=10$.}
\label{fig:sim16}
\end{figure*}

We also tried using a positive adjusted eigenvalue bound (still less than the maximum eigenvalue real-part). This bound still leads to the application of static systemic inhibition, but the resulting maximal eigenvalue real-part of the inhibited system in \eqref{staticinhib} is positive, and not negative as with a negative adjusted eigenvalue bound. The problem is to determine whether there is an adjusted eigenvalue bound $\Lambda^*$ (or equivalently a systemic inhibition parameter $\delta$) for which we will observe sustained phase patterns. As it turns out, for certain choices of $\Lambda^*$, i.i.d. simulations do sometimes, but not always, display sustained phase patterns. This is demonstrated in Figs. \ref{fig:sim15} and \ref{fig:sim16}. Both simulations used coupler A, independently sampled initial phases from a uniform distribution on $[0,2\pi]$, independently sampled initial amplitudes from a uniform distribution on $(0,1]+1000$, used static systemic inhibition with an adjusted eigenvalue bound of $\Lambda^*=10$, and were run independently. We see phase pattern formation, and decreased sample entropy, in Fig \ref{fig:sim15} but not in Fig. \ref{fig:sim16}. Note that the amplitudes in Fig. \ref{fig:sim15} grow large and seemingly without bound similar to previous simulations with pattern formation, whereas the amplitudes in Fig. \ref{fig:sim16} damp to low values similar to previous simulations without pattern formation.

\begin{figure*}
\begin{center}
\resizebox{2\columnwidth}{!}{\includegraphics{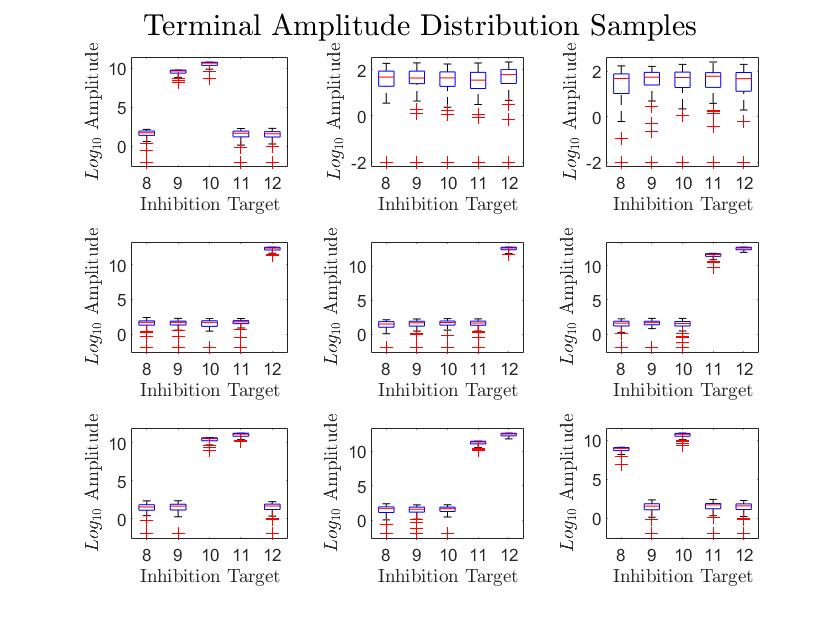}}
\end{center}
\caption{Independent samples of terminal distributions of simulated amplitude processes run for $T=2$ with Mexican Hat coupling \eqref{mexhat} with coupler A; $(b_1,b_2,d_1,d_2)=(1.1,1,5,6.12)$, for various adjusted eigenvalue bounds. The maximum eigenvalue real part of the matrix $\mathcal{L}(\mathbb{C}$), given by \eqref{systemmat}, with these parameters is 0.01516. Initial amplitudes are selected from a uniform distribution on $(0,1]+1000$ and initial phases from a uniform distribution on $[0,2\pi]$. Note that each of the five simulations corresponding to the five box plots in each subplot are Independent as are the simulations between each subplot. Red pluses represent outliers from the boxes.}
\label{fig:boxs1}
\end{figure*}

The results in Figs. \ref{fig:sim15} and \ref{fig:sim16} suggest that both amplitude increase and sustained phase pattern formation occur stochastically, i.e. that pattern formation only occurs with some probability under static systemic inhibition with a small positive real part of the eigenvalues. To explore this further, we performed nine i.i.d. simulations using the same specifications as Figs. \ref{fig:sim15} and \ref{fig:sim16} for a duration of $T=2$, for a range of adjusted eigenvalue bounds $\Lambda^*=8,9,10,11$, and $12$. Box plots of the amplitude distributions at time $t=2$ are given for each simulation in Fig. \ref{fig:boxs1}. For each of the adjusted eigenvalue bounds, the amplitude processes grow to be very large only sometimes. Moreover, it is worthwhile to note that only one of nine simulations have amplitude growth for adjusted eigenvalue bounds of $\Lambda^*=8$ or $\Lambda^*=9$ (when inhibition is set to $\delta=107.8$ and $\delta=106.8$, respectively). By comparison, this occurred for three of nine simulations with adjusted eigenvalue bounds of 10 or 11 ($\delta=105.8$ and $\delta=104.8$, respectively), and for four of 10 simulations with adjusted eigenvalue bounds of 12 ($\delta=103.8$). This suggests that the probability of amplitudes growing large (which has been demonstrated to be associated with pattern formation in the phases) decreases as static systemic inhibition increases.


\subsection{Plastic Systemic Inhibition Permits Phase Patterns with Bounded Amplitudes}

\begin{figure*}
\begin{center}
\resizebox{2\columnwidth}{!}{\includegraphics{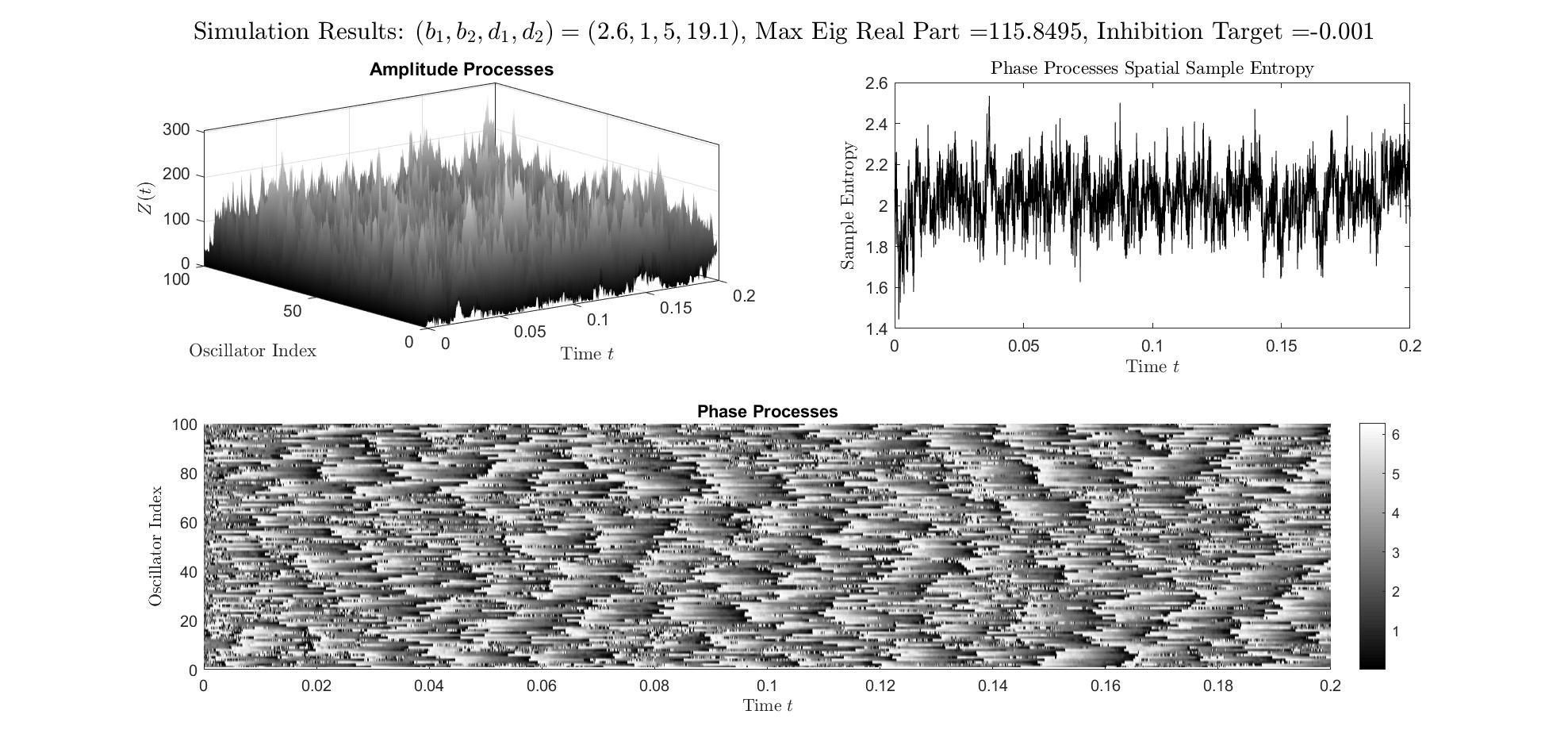}}
\end{center}
\caption{Phase and amplitude process simulation with Mexican Hat coupling \eqref{mexhat} with coupler A; $(b_1,b_2,d_1,d_2)=(2.6,1,5,19.1)$. The maximum eigenvalue real part of the matrix $\mathcal{L}(\mathbb{C}$), given by \eqref{systemmat}, with these parameters was 115.8. Initial amplitudes were selected from a uniform distribution on $(0,1]$ and initial phases from a uniform distribution on $[0,2\pi]$.  Binary-type plastic systemic inhibition \eqref{plasticinhib} was used with $\delta=115.9$ -- corresponding to an adjusted eigenvalue bound of $\Lambda^*=-10^{-3}$ -- with a plastic amplitude threshold of $z^*=100$.}
\label{fig:sim17}
\end{figure*}

\begin{figure*}
\begin{center}
\resizebox{2\columnwidth}{!}{\includegraphics{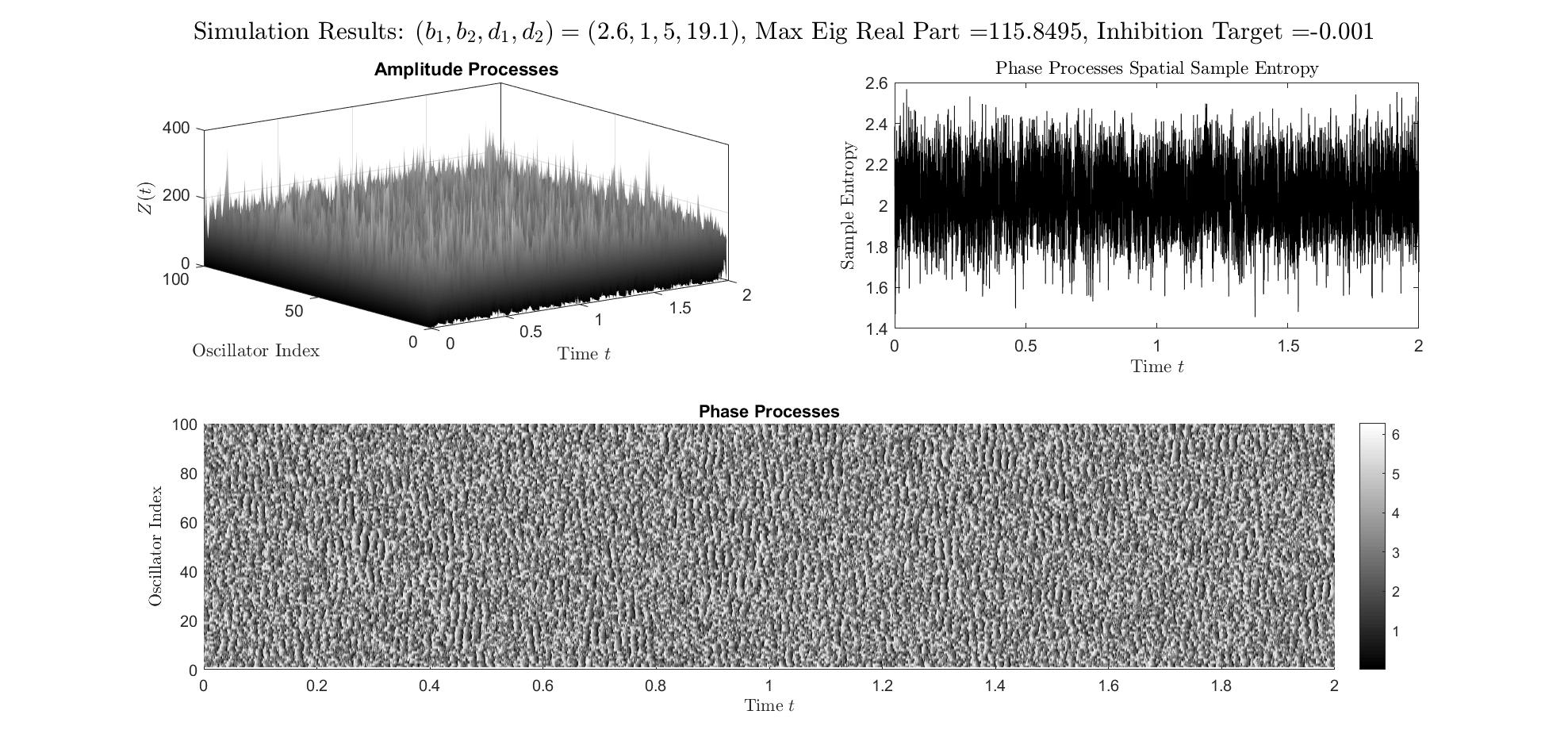}}
\end{center}
\caption{Phase and amplitude process simulation with Mexican Hat coupling \eqref{mexhat} with coupler A; $(b_1,b_2,d_1,d_2)=(2.6,1,5,19.1)$. The maximum eigenvalue real part of the matrix $\mathcal{L}(\mathbb{C}$), given by \eqref{systemmat}, with these parameters was 115.8. Initial amplitudes were selected from a uniform distribution on $(0,1]$ and initial phases from a uniform distribution on $[0,2\pi]$. Saturation-type plastic systemic inhibition \eqref{staticinhib} was used with $\delta=115.9$ -- corresponding to an adjusted eigenvalue bound of $\Lambda^*=-10^{-3}$ -- with a plastic amplitude threshold of $z^*=100$.}
\label{fig:sim20}
\end{figure*}

\begin{figure*}
\begin{center}
\resizebox{2\columnwidth}{!}{\includegraphics{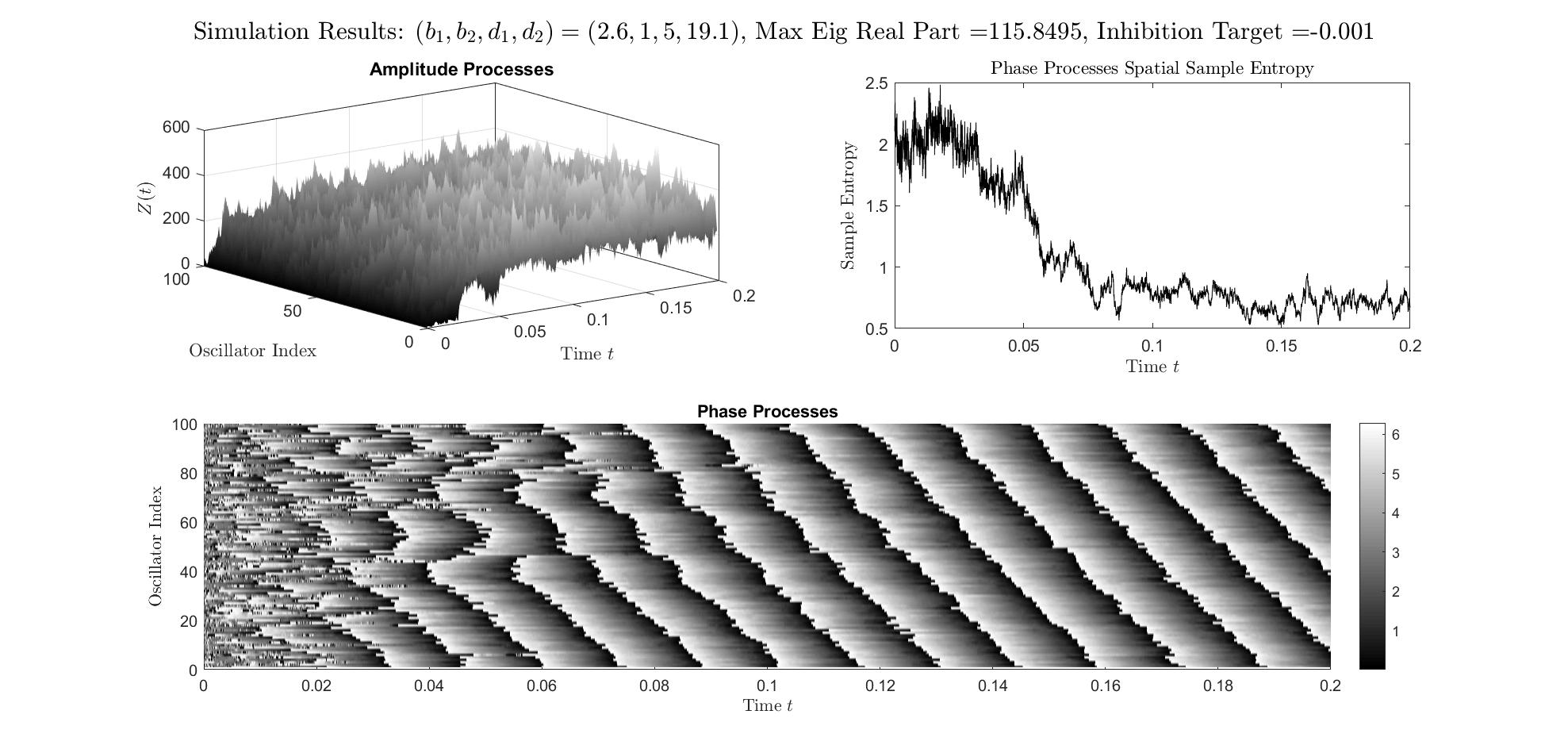}}
\end{center}
\caption{Phase and amplitude process simulation with Mexican Hat coupling \eqref{mexhat} with coupler A; $(b_1,b_2,d_1,d_2)=(2.6,1,5,19.1)$. The maximum eigenvalue real part of the matrix $\mathcal{L}(\mathbb{C}$), given by \eqref{systemmat}, with these parameters was 115.8. Initial amplitudes were selected from a uniform distribution on $(0,1]$ and initial phases from a uniform distribution on $[0,2\pi]$.  Binary-type plastic systemic inhibition \eqref{plasticinhib} was used with $\delta=115.9$ -- corresponding to an adjusted eigenvalue bound of $\Lambda^*=-10^{-3}$ -- with a plastic amplitude threshold of $z^*=300$.}
\label{fig:sim19}
\end{figure*}

\begin{figure*}
\begin{center}
\resizebox{2\columnwidth}{!}{\includegraphics{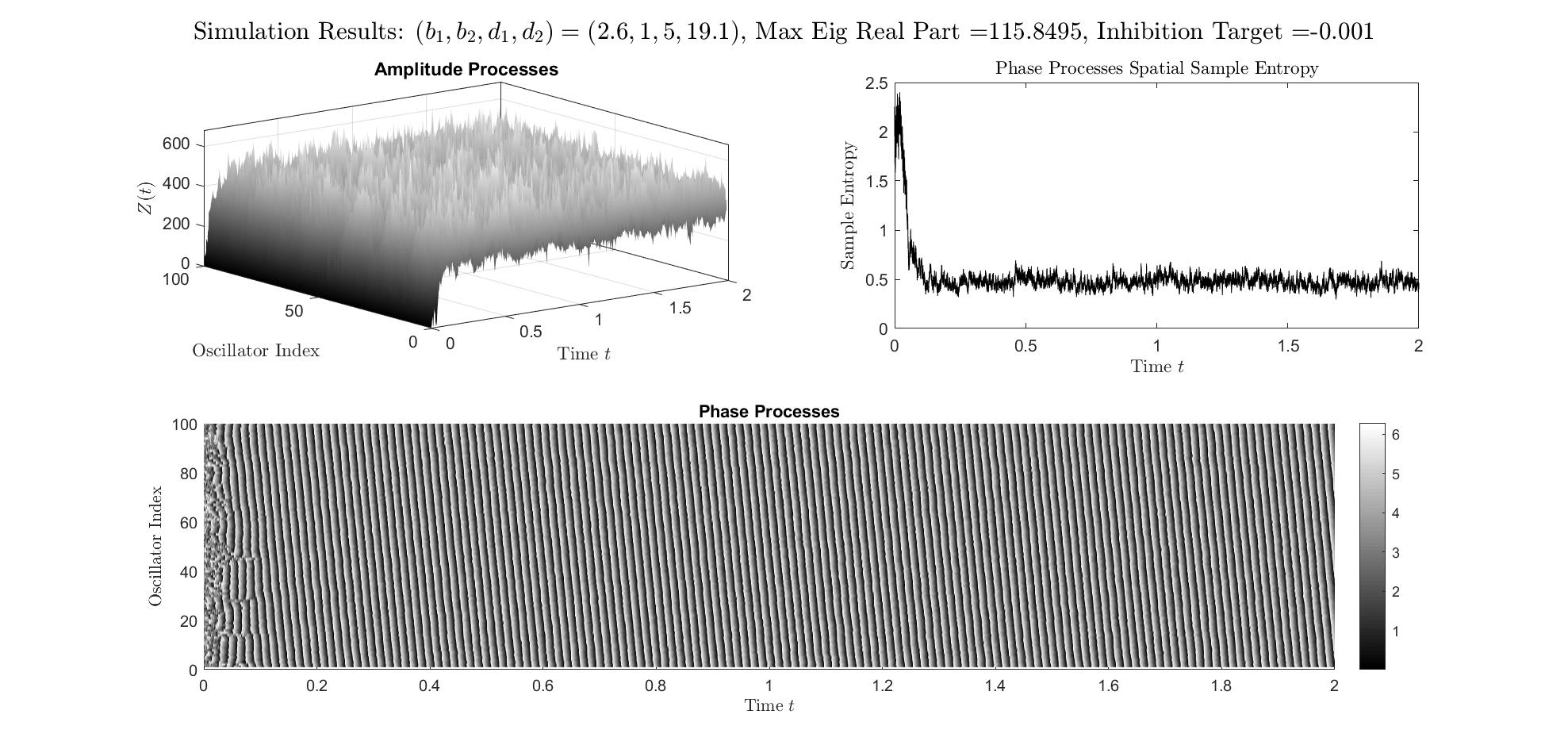}}
\end{center}
\caption{Phase and amplitude process simulation with Mexican Hat coupling \eqref{mexhat} with coupler A; $(b_1,b_2,d_1,d_2)=(2.6,1,5,19.1)$. The maximum eigenvalue real part of the matrix $\mathcal{L}(\mathbb{C}$), given by \eqref{systemmat}, with these parameters was 115.8. Initial amplitudes were selected from a uniform distribution on $(0,1]$ and initial phases from a uniform distribution on $[0,2\pi]$. Saturation-type plastic systemic inhibition \eqref{staticinhib} was used with $\delta=115.9$ -- corresponding to an adjusted eigenvalue bound of $\Lambda^*=-10^{-3}$ -- with a plastic amplitude threshold of $z^*=300$.}
\label{fig:sim21}
\end{figure*}

To study the effects of plastic systemic inhibition, we ran simulations with coupler A and binary-type and saturation-type plastic systemic inhibition (see Section \ref{inhibschemes}). The results are shown in Figs. \ref{fig:sim17} and \ref{fig:sim20} for $z^*=100$ and in Figs. \ref{fig:sim19} and \ref{fig:sim21} for $z^*=300$, for binary-type and saturation-type inhibition, respectively. The result for binary-type plastic inhibition is shown for $z^*=200$ in Fig. S11 -- in which the results are intermediate between Figs. \ref{fig:sim17} and \ref{fig:sim19}. There is only a weak phase pattern apparent in Figs \ref{fig:sim17} and \ref{fig:sim20} ($z^*=100$), with relatively high sample entropy, but Figs. \ref{fig:sim19} and \ref{fig:sim21} ($z^*=300$) display clear phase patterns, with relatively small sample entropies. There is also a clear difference between the amplitudes in these simulations: there are no apparent amplitude patterns, but average amplitudes range from about 100 when $z^*=100$, to 300 when $z^*=300$, as would be expected given how the plastic systemic inhibitory mechanisms were designed. Thus the appearance of phase patterns under these conditions depends on the magnitude of the amplitudes, and does not depend on whether the amplitudes display spatial patterns. Moreover, under these conditions phase patterns can appear in the absence of very large oscillatory amplitudes, making this situation much more similar to actual neural activity in brains. Thus plastic systemic inhibition of the sort implemented here can limit amplitudes to more biological ranges while allowing spatial patterns to form in the phases. 

\section{Discussion}\label{disc}
\subsection{Comparing methods of systemic control}
Beginning with the paper of Wilson and Cowan \cite{WC1972} most studies involving neural population models have used a sigmoid function to bound the values of the variables, which should remain in a dynamic range in order to be biologically meaningful. Such a device has also been extended to neural fields with coupling \cite{FI2015}. However, we were interested in devising a mechanism that achieves the same goal without introducing non-linearity in this way. By finding such a mechanism of bounding variable values, we might preserve the connections that our system has to the literature surrounding quasi-cyles, such as that in \cite{BG2010}, \cite{GMW2015}, \cite{GMW2016}, and \cite{GW2019}. By doing so, it is possible that important questions about neural dynamics could be answered in future research on quasi-cycles. 

Moreover, the saturation-type mechanism we proposed in this paper has a closer connection to the underlying biological system than do the sigmoids in previous works. Both mechanisms operate under the principle that the firing rates of neuron populations reach a point of saturation, however our mechanism can be interpreted as adaptively modifying the system parameters in \eqref{eimatrix1} in response to the overall magnitude of the variables. Sigmoids, on the other hand, simply `correct' the variables, by mapping them to a desired range of values. Bounding variables by a modification of the parameters in \eqref{eimatrix1} is not new; in works such as \cite{AHSJBHBW2018}, systemic inhibition was implemented by modifying the parameter $S_{EI}$. However, this approach only affects the oscillations of each EI-pair, by \eqref{omega}, and not the damping $-\lambda$ of the EI-pairs, as per \eqref{lambda}. By contrast, our systemic inhibitory mechanisms subtract $\delta$ (in the static mechanism) or the diagonals of $\delta \xi(t)$ (in the plastic mechanism) from the diagonals of the matrix $\mathcal{L}(\mathbb{C})$, which are copies of the diagonals of $\mathbb{B}$ containing the intrinsic damping parameter $-\lambda$. Thus, our systemic inhibitory mechanisms are akin to selectively increasing the damping parameter $\lambda$ for each EI-pair individually, while holding all other parameters constant. In particular the saturation-type mechanism subtracts $\delta (1/(1+\max(0,z^*-Z_k))$ from the damping term of each process, thus increasing the damping parameter more the nearer the amplitude is to the saturation threshold, $z^*$, which we conceptualize as the maximum firing rate of a neural population.

For each individual, non-normalized system we have $\lambda=((1-S_{EE})/\tau_E +(1+S_{II})/\tau_I)/2$ (\eqref{lambda}). Hence, in terms of the system parameters $(S_{EE},S_{IE},S_{EI},S_{II})$, changes in $\lambda$ would correspond to changes in self-excitation ($S_{EE}$) and/or in self-inhibition ($S_{II}$). Either a decrease in self-excitation (decrease $S_{EE}$) or an increase in self-inhibition (increase $S_{II}$) would result in an increase in damping by making $-\lambda$ more negative. In our normalized systems it was possible to change $\lambda$ without changing $\omega$ by simply changing the $\lambda$ in the matrix $\mathbb{B}$ and leaving $\omega$ unchanged (via either \eqref{staticinhib} or \eqref{plasticinhib}). But, in the non-normalized original system, \eqref{eimatrix}, in order to maintain the oscillation frequency at the same value when changing $S_{EE}$ or $S_{II}$, a change in one or both of the interaction efficacies ($S_{EI}, S_{IE}$) in \eqref{omega} would be necessary. It would make sense for changes in synapses of the same neuron population to occur at the same time. For example, a decrease in $S_{EE}$ could be accompanied by a corresponding decrease in $S_{EI}$, increasing damping and maintaining the same oscillation frequency. A similar mechanism could be implemented by the inhibitory population, with increases in both $S_{II}$ and $S_{IE}$. These changes in synaptic efficacy would occur rapidly, in an activity-dependent way, similar to the saturation-type plastic inhibition we have studied. Such rapid changes in synaptic efficacies are observed in real neural systems \cite{Kasai2010}, so this does constitute a biologically plausible mechanism for bounding oscillatory amplitudes.

\subsection{Parsing the Sources of Stability}\label{parse}

It is natural to think of our system \eqref{system} as a coupling of the otherwise
i.i.d. Ornstein-Uhlenbeck (OU) processes satisfying \eqref{eimatrix}. A further aim, after our present
results is to find behaviors that generalize over large families of
couplings, which are highly structured but at the same time may be regarded
as `partially random.' Some randomness might allow us to take into account
that the brain structures of individuals develop according to a combination
of tight rules together with some stochasticity.

In order to study separately the nature of neural systems with randomness
of connectivity structure, some authors including \cite{IP2020} \cite{WT2013} \cite{RA2006}
have studied models of the form
\begin{equation}\begin{split}\label{ipeq}
\frac{dx_i}{dt}=-\frac{x_i}{\tau}+\sum_{j=1}^n J_{i,j}S(x_j), \hspace{.25cm}i=1,...,n.
\end{split}\end{equation}
Here $S$ is a sigmoid confining the values of $x_i$ to $[-1, 1]$, $J$ is a
(one-way) connectivity matrix, $\frac{1}{\tau}$ is a decay rate that we can compare to
the $\lambda$ (\eqref{lambda}) in the present paper. The aim of \cite{IP2020} is to construct $J$, which is stochastic with strong structural restrictions in terms of, e.g.,
excitatory-inhibitory balance, in various forms, such that the stability of
the system in terms of the distribution of eigenvalues of $J$ is controlled
with increasing system size, $n$. They find that a critical value of $\frac{1}{\tau}$,
at which the stability of the system changes as the parameter $\tau$
increases, is ($\sigma_{\text{eff}} \sqrt n$). Here $\sigma_{{\text{eff}}}$ is a variance related to
$J$.

In comparing \eqref{ipeq} with \eqref{system} we are representing our OU-type system, \eqref{system}, by just the small, negative part of its eigenvalue pair. The question of pattern formation is absent, but the question of boundedness of the process is present and handled by the sigmoid, $S$, in \eqref{ipeq}. The question arises: suppose we replace $S$ with a ``plastic inhibition", increasing $\frac{1}{\tau}$ when ever $|x_i|$ exceeds a certain value. Do results similar to those of \cite{IP2020} hold?

\section{Conclusions}\label{concl}
In \cite{GW2019}, where we studied synchronization patterns of neural fields of coupled EI-systems in terms of their amplitude and phase processes, we limited consideration to time intervals in which the system remained inside some amplitude bound. There we found that quasi-cycle phases quickly synchronized to form patterns, even with weak coupling, whereas amplitudes formed patterns only at somewhat greater coupling strength. In the present paper we introduced a more realistic bounding device than the sigmoid, which we suggest can be construed as neural plasticity, and which can produce rapid phase pattern formation, while at the same time amplitudes form a bounded, apparently stationary, stochastic field with no apparent patterning. This would seem to be a significant improvement in this direction over previous approaches.

We have demonstrated, via numerical simulations, the important role of amplitude magnitudes in the formation and sustaining of phase patterns for the system given by \eqref{system}. The only way phase patterns were stable in simulations of \eqref{system} was if the amplitudes grew without bound. A static systemic inhibitory mechanism (\eqref{staticinhib}) was used to attempt to bound the amplitudes while sustaining pattern formations. Despite successfully bounding the amplitudes, such an intervention was found to prevent and destroy phase patterns. However, a plastic systemic inhibitory mechanism (\eqref{plasticinhib}) was found to fully resolve the issue -- bounding amplitudes and allowing for sustained phase pattern formation.

Nonetheless, there are still issues to resolve. One issue is that a mechanism must be found in actual neural settings which fulfills the role of a plastic systemic inhibitory mechanism as envisioned in the present paper. The saturation-type mechanism introduced in this paper simply serves as a proof of concept that a plausible biological function in the brain could act as such an inhibitory mechanism, while the binary-type mechanism demonstrates that even a crude mechanism can work. Furthermore, the conjectured connections to \cite{VSZCG2011}, \cite{S2017}, and \cite{HJCL2015} need to be confirmed. Even if a plastic systemic inhibitory mechanism is discovered in the brain that bounds amplitudes and allows phase patterns, it could possibly operate independently of the functions described in \cite{VSZCG2011}, \cite{S2017}, and \cite{HJCL2015}. Another issue to resolve is to explore the mathematics behind the stability in our system with plastic systemic inhibition, as discussed in Section \ref{parse}. Indeed, more work is required to understand exactly why and how plastic systemic inhibition was the key to bounding amplitudes while allowing phase patterns, and whether such a mechanism is present in actual biological contexts. We hope the present paper will stimulate such work and further interest in stochastic neural field models.


\appendix*
\section{Change of Variables with It\^{o}'s formula}

Before carrying out the change of variables for the amplitude and phase processes, we compute the noise terms,
\begin{equation}
\mathbb{E}=\mathbb{Q}^{-1}\mathbb{N}=
\begin{pmatrix}
\frac{-\sigma_E}{\omega\tau_E} & \frac{-1+S_{EE}+\lambda_k\tau_E}{\omega S_{IE}\tau_E\sigma_I}\\
{0} & \frac{\sigma_I}{S_{IE}},
\end{pmatrix}
\end{equation}
so that,
\begin{equation}
\mathbb{E} d \mathbb{W}_k(t)=\begin{pmatrix} a dW_k^E+bdW_k^I\\cdW_k^I \end{pmatrix},
\end{equation}
where,
\begin{equation}\label{a}
a=\frac{-\sigma_E}{\omega\tau_E}
\end{equation}
\begin{equation}\label{b}
b=\frac{-1+S_{EE}+\lambda\tau_E}{\omega S_{IE}\tau_E}\sigma_I,
\end{equation}
and,
\begin{equation}\label{c}
c=\frac{\sigma_I}{S_{IE}}.
\end{equation}
In the calculations that follow, we will see the terms $(du_k)^2$, $(dv_k)^2$, and $du_k \cdot dv_k$, which we will first compute here. In computing each of $(du_k)^2$ and $(dv_k)^2$, there are three terms: one term multiplied by $(dt)^2=0$, a second term multiplied by $dt \cdot (adW_k^E+bdW_k^I)=0$ (resp. $dt \cdot cdW_k^I=0$), and a third term multiplied by $(adW_k^E+bdW_k^I)^2$ (resp. $(cdW_k^I)^2$). These third terms simplify, respectively as,
\begin{equation} \begin{split}
&(adW_k^E+bdW_k^I)^2\\
&=a^2(dW_k^E)^2+b^2(dW_k^I)^2+2abdW_k^EdW_k^I\\
&=(a^2+b^2)dt,
\end{split}\end{equation}
and,
\begin{equation}
(cdW_k^I)^2=c^2 dt.
\end{equation}
Thus 
\begin{equation}\label{usquared}
(du_k)^2=(a^2+b^2)dt
\end{equation}
and,
\begin{equation}\label{vsquared}
(dv_k)^2=c^2dt.
\end{equation}
When computing $du_k \cdot dv_k$, there are four terms: one multiplied by $(dt)^2=0$, another multiplied by $dt \cdot adW_k^E+bdW_k^I=0$, another multiplied by $dt \cdot cdW_k^I=0$, and the fourth term is $(adW_k^E+bdW_k^I) \cdot (cdW_k^I)$, which we expand as,
\begin{equation}\begin{split}
&(adW_k^E+bdW_k^I) \cdot (cdW_k^I)\\
&=acdW_k^E \cdot dW_k^I+bc dW_k^I\cdot dW_k^I\\
&=bc dt,
\end{split}\end{equation}
so that,
\begin{equation}\label{uvprod}
du_k \cdot dv_k=bc  dt.
\end{equation}

Now to change variables we use It\^{o}'s formula, which says that for a smooth function $f: \mathbb{R}^2 \rightarrow \mathbb{R}$, we have,
\begin{equation}
df \begin{pmatrix}u_k \\ v_k \end{pmatrix}=(\nabla f)^\top \begin{pmatrix} du_k\\dv_k \end{pmatrix}+\frac{1}{2} \begin{pmatrix} du_k\\dv_k \end{pmatrix}^\top Hf \begin{pmatrix} du_k\\dv_k \end{pmatrix},
\end{equation}
where $Hf$ is the Hessian matrix,
\begin{equation}
Hf \begin{pmatrix}u_k\\v_k\end{pmatrix}=\begin{pmatrix}\frac{\partial^2 f}{\partial u_k^2} & \frac{\partial^2 f}{\partial u_k \partial v_k}\\
\frac{\partial^2 f}{\partial v_k \partial u_k} & \frac{\partial^2 f}{\partial v_k^2}.
\end{pmatrix}
\end{equation}
\subsection{The Amplitude Process}
For $Z_k \begin{pmatrix}u_k\\v_k \end{pmatrix}=\sqrt{u_k^2+v_k^2}$ we compute,
\begin{equation}
\nabla Z_k \begin{pmatrix}u_k\\v_k \end{pmatrix}=\frac{1}{Z_k}\begin{pmatrix}u_k\\v_k \end{pmatrix},
\end{equation}
and so,
\begin{widetext}
\begin{equation}\begin{split}
&(\nabla Z_k)^\top \begin{pmatrix}du_k\\dv_k \end{pmatrix}=\frac{1}{Z_k}(u_k  du_k+v_k  dv_k)\\
&=\frac{1}{Z_k} \bigg( (-\lambda u_k^2+\omega u_k v_k + u_k \big(\sum_j \mathbb{C}_{k,j}u_j \big) -\omega u_k v_k -\lambda v_k^2+v_k\sum_j \mathbb{C}_{k,j}v_j) \bigg) dt
+\frac{u_k (adW_k^E+bdW_k^I)+ v_k c dW_k^I}{Z_k}\\
&=\bigg(-\lambda Z_k+\frac{1}{Z_k}\sum_j \mathbb{C}_{k,j} (u_ku_j+v_kv_j) \bigg) dt+dR_k,
\end{split}
\end{equation}
where
\begin{equation}
dR_k:=\frac{u_k (adW_k^E+bdW_k^I)+ v_k cdW_k^I}{Z_k}
=a\cos(\theta_k(t)) dW_k^E+ (b\cos(\theta_k(t))+c \sin (\theta_k(t)))dW_k^I.
\end{equation}
Thus,
\begin{equation}
(\nabla Z_k)^\top \begin{pmatrix}du_k\\dv_k \end{pmatrix}=-\lambda Z_k+\sum_j \mathbb{C}_{k,j}Z_j(\cos \theta_k \cos \theta_j  \\+ \sin \theta_k \sin \theta_j) dt+ dR_k.
\end{equation}
Now we compute,
\begin{equation}
\frac{1}{2}H Z_k \begin{pmatrix}u_k\\v_k \end{pmatrix}=\frac{1}{2Z_k} \mathbb{I}_2-\frac{1}{2Z_k^3}\begin{pmatrix}u_k^2 & u_k v_k\\u_k v_k & v_k^2 \end{pmatrix},
\end{equation}
where $\mathbb{I}_2$ is the $2 \times 2$ identity matrix. We have, using \eqref{usquared} and \eqref{vsquared},
\begin{equation}
\begin{pmatrix}u_k \\v_k \end{pmatrix}^\top (\frac{1}{2Z_k}\mathbb{I}_2)\begin{pmatrix}u_k \\ v_k \end{pmatrix}=\frac{1}{2Z_k}\big( (du_k)^2+(dv_k)^2 \big)=\frac{a^2+b^2+c^2}{2Z_k}dt.
\end{equation}
Now we compute,
\begin{equation}
\frac{1}{2Z_k^3} \begin{pmatrix}du_k\\dv_k\end{pmatrix}^\top \begin{pmatrix}u_k^2 & u_k v_k\\u_k v_k & v_k^2 \end{pmatrix} \begin{pmatrix} du_k\\dv_k \end{pmatrix}=\frac{1}{2Z_k^3}(u_k^2(du_k)^2+v_k^2(dv_k)^2 + 2u_kv_k du_k \cdot dv_k).
\end{equation}
Using \eqref{usquared}, \eqref{vsquared}, and \eqref{uvprod}, we have that,
\begin{equation} \label{hess}
\begin{split}
&\frac{1}{2Z_k^3} \begin{pmatrix}du_k\\dv_k \end{pmatrix}^\top \begin{pmatrix}u_k^2 & u_k v_k\\u_k v_k & v_k^2 \end{pmatrix} \begin{pmatrix} du_k \\ dv_k \end{pmatrix}= \frac{(a^2+b^2)u_k^2+c^2v_k^2+2u_kv_kbc}{2Z_k^3}dt\\
&=\frac{(a^2+b^2) \cos (\theta_k(t))^2+c^2 \sin (\theta_k(t))^2+bc\sin(2\theta_k(t))}{2Z_k}dt.
\end{split}\end{equation}
So, using the substitution $\cos(\theta_k-\theta_j)=\cos \theta_k \cos \theta_j + \sin \theta_k \sin \theta_j$ the amplitude process is given by,
\footnotesize
\begin{equation}\label{AZ}
dZ_k=\bigg( \frac{a^2+b^2+c^2-(a^2+b^2) \cos (\theta_k(t))^2-c^2 \sin (\theta_k(t))^2-bc\sin(2\theta_k(t))}{2Z_k}-\lambda Z_k
+\sum_j \mathbb{C}_{k,j}Z_j\cos(\theta_k-\theta_j) \bigg) dt+dR_k.
\end{equation}

\subsection{The Phase Process}
For $\theta_k=\arctan \frac{v_k}{u_k}$ we compute,
\begin{equation}
\nabla \theta_k=\frac{1}{Z_k^2}\begin{pmatrix}-v_k\\u_k \end{pmatrix},
\end{equation}
and so,
\begin{equation}
(\nabla \theta_k)^\top \begin{pmatrix} du_k\\dv_k \end{pmatrix}= \frac{1}{Z_k^2}\bigg(\lambda v_ku_k-\omega v_k^2-\omega u_k^2-\lambda u_k v_k +\sum_j \mathbb{C}_{k,j}(-v_ku_j+u_kv_j) \bigg) dt+dS_k(t),
\end{equation}
where,
\begin{equation}
dS_k(t):=\frac{-v_k(adW_k^E+bdW_k^I)+u_kcdW_k^I}{Z_k^2}=\frac{-a\sin(\theta_k(t))dW_k^E+(c\cos(\theta_k(t))-b\sin(\theta_k(t)))dW_k^I}{Z_k}.
\end{equation}
And so,
\begin{equation}
(\nabla \theta_k)^\top \begin{pmatrix} du_k\\dv_k \end{pmatrix}= \bigg(-\omega+\sum_j \mathbb{C}_{k,j} \frac{Z_j}{Z_k}(- \sin \theta_k \cos \theta_j+\cos \theta_k \sin \theta_j)\bigg) dt+dS_k(t).
\end{equation}
Now we compute the Hessian,
\begin{equation}
H \theta_k=\frac{1}{Z_k^2} \begin{pmatrix}\frac{2u_k v_k}{Z_k^2} & 1-\frac{2u_k^2}{Z_k^2}\\
-1+\frac{2v_k^2}{Z_k^2} & \frac{-2u_kv_k}{Z_k^2} ,\end{pmatrix}
\end{equation}
and find, using \eqref{uvprod},
\begin{equation}\begin{split}
&\frac{1}{2}\begin{pmatrix} du_k\\dv_k \end{pmatrix}^\top H\theta_k\begin{pmatrix}du_k\\dv_k\end{pmatrix}\\
&=\frac{1}{2Z_k^2} \Bigg(((du_k)^2+(dv_k)^2)\big[ \frac{2u_kv_k}{Z_k^2}+\frac{-2u_kv_k}{Z_k^2} \big] dt+du_kdv_k \big[ 1-\frac{2u_k^2}{Z_k^2}-1+\frac{2v_k^2}{Z_k^2} \big] \Bigg)\\
&=bc\frac{v_k^2-u_k^2}{Z_k^4}dt=bc\frac{1-2\cos(\theta_k(t))^2}{Z_k^2}dt.
\end{split}\end{equation}
And so, using the substitution $\sin(\theta_j-\theta_k)=- \sin \theta_k \cos \theta_j+ \cos \theta_k \sin \theta_j$ the phase process is given by,
\begin{equation}\label{Atheta}
d\theta_k=\bigg(bc\frac{1-2\cos(\theta_k(t))^2}{Z_k^2}-\omega+\sum_j \mathbb{C}_{k,j} \frac{Z_j}{Z_k}\sin(\theta_j-\theta_k)\bigg) dt+dS_k.
\end{equation}
\end{widetext}

\subsection{Differences between It\^{o} Transformations}
The amplitude and phase processes given in \eqref{AZ} \eqref{zeq} and \eqref{Atheta} \eqref{thetaeq} are different from those derived in \cite{GW2019}. This arises from different treatments of lower order terms and of the Brownian motions $dW_k^E$ and $dW_k^I$. Here we retained lower order terms whereas in \cite{GW2019} they were dropped. Initially in \cite{GW2019} and here the noise terms are expressed in vector form as $\mathbb{E} d\mathbb{W}_k$, where the matrix $\mathbb{E}$ is upper-triangular. But in the It\^o transformation in \cite{GW2019}, to make the derivation simpler, it was assumed that $\mathbb{E}=\mathbb{I}$, and thus the stochastic terms had coefficients of 1 (i.e. $a=c=1$ from \eqref{acoef} and \eqref{ccoef}, whereas here we retained the coefficient $b$ \eqref{bcoef}, and did not set $a=c=1$. We can account for the discrepancies between the It\^{o} derivation of \cite{GW2019} and the present one by interpreting the noises in \cite{GW2019} as being independent with equal coefficients \textit{after} the change of basis via the matrix $\mathbb{Q}$ \eqref{matrixq} has taken place, instead of being independent in the original system (i.e. \eqref{eimatrix}).

It is simple to verify that the deterministic term in \eqref{AZ} simplifies to the equation of the coupled amplitude processes in \cite{GW2019} in the case where we set $a=c=1$. The noise term $dR_k$ in \eqref{AZ} can be identified as a Brownian motion with coefficient 1. The details are complicated but the main idea is that the increments of $R_k$ are given by,
\begin{equation}\begin{split}
&R_k(t+\Delta t)-R(t)\\
&=\int_t^{t+\Delta t} \frac{u_k(t)}{Z_k(t)} dW_k^E(t)+\int_t^{t+ \Delta t} \frac{v_k(t)}{Z_k(t)} dW_k^I(t),
\end{split}\end{equation}
where the stochastic integrals can be written as the limits of the sums,
\begin{equation}\label{integralform}
\begin{split}
&\sum_i \frac{u_k(t_i)}{Z_k(t_i)}\big(W_k^E(t_{i+1}) - W_k^E(t_{i}) \big)\\
&+\sum_i \frac{v_k(t_i)}{Z_k(t_i)}\big(W_k^I(t_{i+1}) - W_k^I(t_{i}) \big)
\end{split}\end{equation}
where the mesh $\{t_i\}$ increases to the mesh $\mathbb{Q} \cap [t, t+\Delta t]$. Noting that the increments in $W_k^E$ and $W_k^I$ in \eqref{integralform} are both $\mathcal{N}(0,t_{i+1}-t_i)$ (and are independent by assumption), we have,
\begin{equation}\begin{split}
&\frac{u_k(t_i)}{Z_k(t_i)}\big(W_k^E(t_{i+1}) - W_k^E(t_{i}) \big)\\
&+ \frac{v_k(t_i)}{Z_k(t_i)}\big(W_k^I(t_{i+1}) - W_k^I(t_{i}) \big)\\
& \sim \mathcal{N}(0,\frac{u_k^2+v_k^2}{Z_k}(t_{i+1}-t_i))\\
&=_d \mathcal{N}(0,t_{i+1}-t_i).
\end{split}\end{equation}
Hence in the limit $\{t_i\} \rightarrow \mathbb{Q} \cap[t, t+\Delta t]$ \eqref{integralform} has a $\mathcal{N}(0,\Delta t)$ distribution, and so it is easy to see that $R_k$ is actually a Brownian motion. Similar calculations show that taking $a=c=1$ and $b=0$ renders \eqref{Atheta} equivalent to the phase process equation in \cite{GW2019}.
\section*{Competing interests}
The authors declare that they have no competing interests.
\section*{Authors' contributions}
All authors contributed to the conceptualization and writing of the paper. The numerical simulations were accomplished by CLM.
\section*{acknowledgements}
This research was supported by grants from the Natural Sciences and Engineering Research Council (NSERC) of Canada to Yaniv Plan (via support for CLM), and to LMW (A9958). We would like to thank Yaniv Plan for contributing funding despite not being directly affiliated with this paper.
%

\end{document}